\documentclass[a4paper,11pt]{article}
\usepackage{jheppub}
\usepackage{lineno}
\usepackage{graphicx}
\usepackage{lipsum}
\usepackage{slashed}
\usepackage{amsmath}
\usepackage[normalem]{ulem}
\usepackage{enumitem}\usepackage{mathtools}
\usepackage{bm}
\usepackage[all]{hypcap}
\usepackage{tikz}
\usepackage{tikz-feynman}
\usepackage{bbm}
\usepackage{xcolor}
\usepackage{subcaption}
\usepackage[font=footnotesize,labelfont=bf]{caption}
\usepackage{mathabx}
\usepackage{pifont}
\usepackage{multirow}
\usepackage{tablefootnote}
\usepackage{cleveref}
\usepackage{xcolor}

\newcommand{\cC}{\mathcal{C}}
\newcommand{\Op}{\mathcal{O}}

\makeatletter\g@addto@macro\bfseries\boldmath\makeatother

\title{
Yukawa coupling deviations from four-quark interactions}

\author{Sophie Renner}
\emailAdd{sophie.renner@glasgow.ac.uk}

\author{and Benjamin Smith}
\emailAdd{b.smith.4@research.gla.ac.uk}

\affiliation{School of Physics and Astronomy, University of Glasgow, Glasgow G12 8QQ, United Kingdom}

\abstract{
New physics generating four-quark interactions can be difficult to detect at the LHC, especially if it does not couple to first generation quarks. We investigate the possible role of such interactions in generating measurable deviations in Higgs-quark couplings. We find important constraints from flavour physics and top-pair production, but show that Higgs-charm and Higgs-bottom couplings remain the most sensitive LHC observables to some four-quark operators involving scalar currents. At FCC-ee, more operators and flavour structures can be tested. We further investigate simple single-particle models that can generate these interactions, including a second Higgs doublet, and scalar and vector diquarks. We find that in some cases, including new scalars most strongly coupled to the third generation of quarks, deviations in Yukawa couplings could offer the first signs of new physics generating four-quark interactions.}

\begin{document}

\maketitle
\flushbottom

\section{Introduction}
Many of the Higgs boson's properties and couplings are now well measured, with the results in broad agreement with expectations within the Standard Model (SM) \cite{ATLAS:2022vkf,CMS:2022dwd}. An exception to this precise picture are the Higgs couplings to the first and second generations of fermions, which remain relatively unconstrained.

In the leptonic sector, the key difficulty is the tiny size of the couplings, nonetheless, there is evidence for the muon Yukawa coupling at the LHC \cite{ATLAS:2025coj} and prospects for observation at HL-LHC \cite{ATLAS:2025eii}. At FCC-ee, a proposed Higgs pole run would enable sensitivity to the electron Yukawa within $O(1)$ of the SM value~\cite{Greco:2016izi,dEnterria:2021xij,FCC:2025lpp,Boughezal:2024yjk,Fatehi:2026gnt}.

In the quark sector, the additional challenge of final-state tagging of Higgs decays means the observation of a SM-like second-generation coupling is not expected until the FCC-ee becomes operational. The third-generation bottom-quark Yukawa coupling provides a benchmark to which other sensitivities can be compared, with combinations of ATLAS and CMS measurements achieving $10$--$20$\% sensitivity \cite{ATLAS:2022vkf, CMS:2022dwd} and a decrease to $3.6$\% expected from the HL-LHC \cite{ATLAS:2025eii}.
Searches for Higgs decays to charm quark-antiquark pairs, produced alongside $W$ or $Z$ bosons \cite{ATLAS:2024yzu, CMS:2022fxs} or a $t\bar t$ pair \cite{CMS:2025dsh} give constraints on the charm Yukawa coupling, with the $t \bar t h$ analysis giving the strongest current bound $|\kappa_c|<3.5$ at 95\% CL (where $\kappa_c\equiv\frac{y_c}{y^{\text{SM}}_c}$). The dominant gluon-fusion production mode has much weaker sensitivity due to high Quantum Chromodynamics (QCD) multijet backgrounds \cite{CMS:2022psv}. Other probes of this coupling include exclusive radiative Higgs decays to vector mesons \cite{Perez:2015lra, Bodwin:2013gca, ATLAS:2022rej, CMS:2024hhg}, associated charm production \cite{Brivio:2015fxa, Dong:2024gts, ATLAS:2024ext, CMS:2025qmm}, $h+\gamma$ production \cite{Aguilar-Saavedra:2020rgo} and modifications to the $p_T$ spectrum of the Higgs \cite{Bishara:2016jga, CMS:2018gwt, ATLAS:2022fnp, CMS:2023gjz}. At HL-LHC, improvements in $c$-tagging techniques are expected to enable a direct search constraint $|\kappa_c|<1.5$ \cite{ATLAS:2025eii, Cepeda:2019klc}.

Constraints on all first- and second-generation couplings, under the assumption of SM third-generation couplings, have been placed from the inclusive production of $h\to 4 \ell$ \cite{CMS:2025xkn}, with the relatively loose bounds representing the strongest current handle on Higgs couplings to $s$, $d$ and $u$ quarks. Global fits assuming HL-LHC sensitivity indicate the untagged branching ratio of the Higgs will provide a leading constraint on the strange Yukawa modifier $|\kappa_s|<13$ \cite{deBlas:2019rxi}, while the strongest HL-LHC projected constraints on $\kappa_{u,d}$ are from off-shell Higgs production \cite{Balzani:2023jas}. 
At FCC-ee, improvements in jet-tagging algorithms and increased computing power will allow sub-percent level measurements of $\kappa_{b,c}$ and a vastly improved constraint on $\kappa_s$ \cite{FCC:2025lpp}. The projected limits on Higgs branching ratios to up and down quarks given in \cite{del_vecchio_2025_3jjdh-6fz97, Selvaggi:2025kmd} can be interpreted as the limits $|\kappa_u|<120$ and $|\kappa_d|<50$ representing an improvement by a factor $2$--$3$. Study of dihadron fragmentation may lead to further improvements in these bounds \cite{Cao:2025wfg}. All the current bounds and projected sensitivities mentioned here are summarised in Tab.~\ref{tab:kappaprojections}.

In the SM, a single parameter determines both the mass of each fermion and its Yukawa coupling to the physical Higgs boson. Hence, knowing the masses precisely, we do not learn anything new from measurements of these couplings within the SM itself. But this degeneracy may be broken by new physics, such that a measured deviation in the Yukawa couplings away from their SM values could be the first sign of new physics. 
Models which can induce significant tree-level contributions to effective light-quark ($c,s,u,d$) Yukawa couplings include two Higgs doublet models (2HDMs)~\cite{Altmannshofer:2016zrn, Ghosh:2015gpa, Giannakopoulou:2024unn} and models containing vector-like quarks (VLQs)~\cite{Erdelyi:2024sls}. In the 2HDM setups, the couplings of new scalars to light fermions lead to key signatures in resonance searches: $\mu^+ \mu^-$ and low mass dijet final states for the model of Ref.~\cite{Altmannshofer:2016zrn}, and $hh$ and $Zh$ final states for the model of Ref.~\cite{Giannakopoulou:2024unn}. For sufficiently massive new doublets, however, large enhancements of light quark Yukawas are possible. For VLQ pairs, avoiding large flavour-changing effects (meson mixing) forces couplings to only one light-generation. Current fits to Higgs data and electroweak precision observables (EWPOs) allow for deviations in $\kappa_{c,s,u,d}$ which would be accessible at HL-LHC. In contrast, an FCC-ee Tera-Z run gives indirect electroweak constraints which are competitive/stronger than the FCC-ee sensitivity to $\kappa_{c,s,u,d}$ \cite{Erdelyi:2024sls}.

\begin{table}[t]
    \centering
    \begin{tabular}{|c|c|c|c|}
          \hline Coupling modifier & Current & HL-LHC & FCC-ee \\ \hline
         $\kappa_u$ & $[-2.4, 2.4]\times10^3$ \cite{CMS:2025xkn} & $[-260,260]$ \cite{Balzani:2023jas} & $[-120, 120]$ \cite{Selvaggi:2025kmd} \\
         $\kappa_c$ & $[-3.5,3.5]$ \cite{CMS:2025dsh} & $[-1.5,1.5]$ \cite{ATLAS:2025eii} & $\pm0.014$ \cite{FCC:2025lpp}\\
         $\kappa_t$ & $0.95\pm0.14$ \cite{ATLAS:2022vkf} & $\pm0.064$ \cite{FCC:2025lpp}& $\pm0.062$ \cite{FCC:2025lpp} \\ \hline
         $\kappa_d$ & $[-9.4, 9.3]\times10^2$ \cite{CMS:2025xkn} & $[-156,156]$ \cite{Balzani:2023jas} & $[-50,50]$ \cite{Selvaggi:2025kmd} \\
         $\kappa_s$ & $[-41,40]$ \cite{CMS:2025xkn} & $[-13,13]$ \cite{deBlas:2019rxi}& $^{+0.77}_{-1.15}$ \cite{FCC:2025lpp} \\
         $\kappa_b$ & $0.90\pm0.22$ \cite{CMS:2023gjz} & $\pm0.072$ \cite{ATLAS:2025eii}& $\pm0.0098$ \cite{FCC:2025lpp} \\ \hline
    \end{tabular}
    \caption{Current experimental constraints and projected sensitivities for $\kappa_q$ at 95\% CL. Where necessary, 68\% constraints have been doubled to estimate the 95\% sensitivity.}
    \label{tab:kappaprojections}
\end{table}

If new physics exists around the TeV scale, it is natural to guess that it could be most strongly connected to the third generation of quarks. This is motivated by solutions to the hierarchy problem which usually predict new top-partner states to cancel the dominant effect of top loops on the Higgs mass, solutions to the SM flavour problem which must explain the much larger masses of the third-generation fermions, and for the phenomenological reason that flavour-universal new physics is already strongly constrained by LHC measurements (see e.g.~\cite{Allwicher:2023shc}). Given the large Yukawa coupling between the Higgs and the top, it is possible that some novel top-containing interactions could first appear indirectly in measurements of Higgs properties.

In this work, we explore the idea of new physics which generates light-quark Yukawa deviations at loop level, via new four-quark interactions involving top quarks. In fact, these interactions provide the \emph{only} viable option for significant loop level contributions to quark Yukawas, meaning that our analysis complements tree level models already covered in the literature, to cover most of the rest of the BSM space that can be tested by measurements of $\kappa_q$. The relevant four-quark interactions necessarily contain chiral-symmetry-breaking \emph{scalar} currents of tops and light quarks, and hence do not appear within commonly-studied subsets of effective operators obeying certain flavour symmetries such as $U(3)^5$ or $U(2)^5$. Our analysis therefore involves new studies of their phenomenology and bounds on their coefficients. We calculate their effects in flavour-changing processes such as meson mixing and top pair production, to determine under what conditions they could first appear in light-quark Yukawa deviations. 
We also explore simple UV completions of these operators.

The paper is structured as follows. We begin in Section~\ref{sec:eft} with a study of quark Yukawa modifications within the Standard Model Effective Field Theory (SMEFT), where additional constraints on the relevant operators limit the $\kappa_q$ deviations that they can generate. We find operators that can induce deviations in $\kappa_c$ or $\kappa_b$ which would be observable at HL-LHC, whereas four-quark operators generating $\kappa_{u,d,s}$ are already constrained by $t\bar{t}$ production and flavour physics, precluding HL-LHC-observable deviations in these couplings from four-quark operators (though in some cases FCC-ee has sufficient sensitivity). Section~\ref{sec:uv_models} explores simple one-particle models that can underlie the four-quark operators which induce $\kappa_c$ or $\kappa_b$, identifying colour-charged vector bosons and a second Higgs doublet as promising options, but which are subject to constraints from direct searches and flavour physics.
We summarise our findings and conclude in Section~\ref{sec:concs}.

\section{Quark Yukawa modifications in the SMEFT}
\label{sec:eft}

We begin with an overview of quark Yukawa modifications in the SMEFT framework.
To operator dimension six, the SMEFT Lagrangian is written 
 \begin{equation}
\mathcal{L_{\text{SMEFT}}}=\mathcal{L}_{\text{SM}}+\sum_{k} \mathcal{C}_k\mathcal{O}_k,
\end{equation}
where $\Op_i$ are local effective dimension six operators and $\cC_i$ are corresponding Wilson coefficients with dimensions TeV$^{-2}$. In the Warsaw basis, the leading order corrections\footnote{Additional corrections to quark Yukawas can arise from $\Op_{H}$, $\Op_{H\Box}$ and $\Op_{HD}$. However, these modifications are both flavour universal and proportional to the corresponding SM Yukawa coupling. Their effects would be most visible in the Yukawa couplings measured with the highest relative precision (and/or in other Higgs couplings or electroweak precision tests). We neglect their effects here since we focus on operators which would first show up in the poorly-constrained first- and second-generation quark Yukawa couplings.} to up- and down-type quark Yukawa matrices are from the operator structures
\begin{equation}
    \mathcal{O}_{uH} =  |H|^2\overline{q}\tilde{H}u , ~~~ \mathcal{O}_{dH} = |H|^2\overline{q}Hd,
\label{eq:Yukd6}
\end{equation}
respectively. The couplings between the physical Higgs boson and the fermions can be written as
\begin{equation}
    \mathcal{L}_{h\bar f f}= -[g_{hff}]_{ij}h\bar f_R^i f_L^j + h.c.,
\end{equation}
where with the inclusion of the effective operators, the coupling matrices are given by 
\begin{equation}
    [g_{hff}]_{ij} = \frac{1}{v}[M_f]_{ij} - \frac{v^2}{\sqrt{2}}[\cC^*_{fH}]_{ji},
\end{equation}
where $f\in{u,d}$. We can diagonalise the mass matrices with biunitary transformations, however, depending on the initial flavour structure of $[\cC^*_{fH}]_{ji}$, $[g_{hff}]_{ij}$ may not be diagonal in flavour space.

The ratio of a quark Yukawa coupling to its SM value is given by
\begin{equation}
    [\kappa_u]_i = 1- \frac{v^3}{\sqrt{2} [m_u]_i}[\hat{\cC}_{uH}]_{ii},~~~ [\kappa_d]_i = 1- \frac{v^3}{\sqrt{2} [m_d]_i}[\check{\cC}_{dH}]_{ii},
    \label{eq:kappa_formulae}
\end{equation}
with SMEFT coefficients evaluated at the electroweak scale ($\mu_{\text{EW}}\approx m_h$), and where we are assuming purely real-valued $[\cC_{qH}]_{ii}$. In calculations of $\kappa_q$, we need to be clear on what we mean by the SM value of the Yukawa couplings, or in other words the definition of $[m_{u,d}]_i$ in the above equations. Here we take the quark masses to be constant non-running values\footnote{Specifically, we take $m_d=4.7\times 10^{-3}$, $m_u=2.2\times 10^{-3}$, $m_s=95 \times 10^{-3}$, $m_c=1.27$, $m_b=4.18$, $m_t=172.6$, all in GeV.} (consistent with other phenomenological work e.g.~\cite{Erdelyi:2024sls}). Since the light quark masses do not play any other role in our calculations (their effects are negligible in amplitudes and kinematics), a different choice here would result in a simple scaling of the relationship between $\cC_{qH}$ and $\kappa_q$. Within the SMEFT, the flavour basis must be specified when labelling flavour indices for the quark doublets. The hatted notation $\hat{\cC}$ indicates that the quark-doublet flavour indices are defined in the basis where the up-type Yukawa matrix is diagonal, while the inverse notation $\check{\cC}$ indicates that they are defined in the basis where the down-type Yukawa matrix is diagonal. Current experimental bounds and projected future sensitivities on all $\kappa_q$ are listed in Tab.~\ref{tab:kappaprojections}.

\subsection{Flavour alignment and indirect constraints}
While the diagonal entries of $\hat{\cC}_{uH}$ and $\check{\cC}_{dH}$ modify flavour-conserving couplings from their SM values, the off-diagonal entries give tree-level contributions to quark-flavour violating processes involving Higgs bosons, which do not appear in the SM. For instance, $[\hat{\cC}_{uH}]_{32/23}$ and $[\hat{\cC}_{uH}]_{31/13}$ modify the branching ratios $\mathcal{B}(t\to hc)$ and $\mathcal{B}(t\to hu)$, which are constrained by the LHC \cite{ATLAS:2024mih}. However, the LHC has little direct sensitivity to flavour-changing Higgs decays to other quark pairs, such as $h\to bs, bd$ \cite{Blankenburg:2012ex}. At FCC-ee, improved flavour tagging and analysis techniques may allow for meaningful direct limits on these processes \cite{Kamenik:2023hvi}.

Indirect constraints on the off-diagonal entries can be derived from the neutral meson mixing observables $\Delta m_D$, $\Delta m_K$, $\Delta m_{B_d}$ and $\Delta m_{B_s}$ \cite{Harnik:2012pb, Blankenburg:2012ex}. These provide the strongest constraints on $[\hat{\cC}_{uH}]_{12/21}$ and all of the off-diagonal entries of $\check{\cC}_{dH}$. We calculate these constraints using the bounds on low-energy Wilson coefficients from \cite{utfit}, where their analysis is based on \cite{UTfit:2007eik}, as proxies for the relevant experimental observables. By the time of FCC-ee, improvements in the determination of the Cabibbo–Kobayashi–Maskawa (CKM) matrix element $V_{cb}$, driven by progress in lattice QCD alongside precision measurements of on-shell $W$ decays to $b$- and $c$-jets at FCC-ee \cite{Marzocca:2024mkc, Liang:2024hox}, are expected to yield a roughly factor of three increase in the new physics scale associated with low-energy coefficients relevant for $B_d$- and $B_s$-mixing \cite{deBlas:2025gyz}. We use this improvement to project bounds on the $[\check{\cC}_{dH}]_{31/13}$ and $[\check{\cC}_{dH}]_{32/23}$ entries. Relevant formulae are summarised in App.~\ref{app:meson_mixing}.

Including also the diagonal components (from Tab.~\ref{tab:kappaprojections}), we present the current symmetrised 95\% CL constraints on all the entries of $\hat{\cC}_{uH}$ and $\check{\cC}_{dH}$, alongside FCC-ee projections where they exist\footnote{In principle, FCC-ee should improve the limits on $t\to h c$ and $t\to h u$, but no projections exist for these decays. Ref.~\cite{Ai:2024nmn} argues that CEPC, with $6\times 10^5$ $t\bar{t}$ pairs, should be sensitive to branching ratios of the order of $10^{-5}$, an improvement of a factor of roughly 2 on current bounds. The current run plan for FCC-ee anticipates $2\times 10^6$ $t\bar{t}$ pairs \cite{FCC:2025lpp}, so may be naively expected to achieve an improvement of a factor of $2-3$ on the corresponding Wilson coefficients with respect to current bounds.}
\[
|\hat{\cC}_{uH}(m_h)| \times \mathrm{TeV}^2 \lesssim
\left\{
\begin{array}{c}
\left(
\begin{array}{ccc}
0.50 & 3.3\times10^{-4} & 0.68\\
 & 0.30 & 0.78\\
 & & 3.1
\end{array}
\right){(\text{current})}, \\[1.5em]\left(\begin{array}{ccc}
2.5\times10^{-2} &  & \\
 & 1.7\times 10^{-3} &\\
 & & 1.0
\end{array}
\right){(\text{FCC-ee}),}
\\[1.5em]
\end{array}
\right.\]
\\[2em]
\[
|\check{\cC}_{dH}(m_h)| \times \mathrm{TeV}^2 \lesssim
\left\{
\begin{array}{c}
\left(
\begin{array}{ccc}
0.42 & 1.1\times10^{-4} & 7.3\times10^{-4}\\
 & 0.35 & 3.2\times10^{-3}\\
 & & 0.13
\end{array}
\right){(\text{current})},\\[1.5em]
\left(
\begin{array}{ccc}
2.2\times10^{-2} &  & 2.5\times10^{-4}\\
 & 1.9\times10^{-2} & 1.0\times10^{-3}\\
 & & 3.9\times10^{-3}
\end{array}
\right){(\text{FCC-ee})},
\\[1.5em]
\end{array}
\right.
\]
where constraints on $[\hat{\cC}_{uH}]_{32/23}$ and $[\hat{\cC}_{uH}]_{31/13}$ are from \cite{ATLAS:2024mih}.
It is clear that, in both sectors, some degree of diagonal alignment is required for FCC-ee-level Yukawa deviations to be consistent with existing $\Delta F=2$ constraints. This requirement is stronger in the down sector where these indirect constraints apply to all off-diagonal entries. In the up sector, the weak bounds on flavour-changing top decays relax the alignment requirement in the third row/column.

If we assume that the UV features some mechanism of alignment, such as Spontaneous Flavour Violation \cite{Egana-Ugrinovic:2018znw}, one may then ask whether this alignment is radiatively stable, given a separation between the new physics scale and $m_h$. In the down sector, the radiative misalignment has a contribution proportional to the large top Yukawa coupling. For instance, the renormalisation group equation (RGE) for $[\check{\cC}_{dH}]_{12}$ contains the term
\begin{equation}
    [\dot{\check{\cC}}_{dH}]_{12} \supset -\frac{3y_t^2}{2(16\pi^2)}V^*_{td}V_{ts} [\check{\cC}_{dH}]_{22},
\end{equation}
which introduces a small misalignment dependent on the new physics scale. Assuming only $[\check{\cC}_{dH}]_{22}$ is non zero at three benchmark UV scales $\Lambda=1,\, 3$ and $10$~TeV, we solve the relevant RGEs using \texttt{wilson} \cite{Aebischer:2018bkb}, thereby also including the additional effect of re-diagonalising the down-type Yukawa matrix at the EW scale \cite{Aebischer:2020lsx}.
The strongest constraint on the resulting radiative contribution to $[\check{\cC}_{dH}]_{12/21}$ is from indirect CP-violation in the $K^0$--$\overline{K}^0$ system, encoded in the parameter $\varepsilon_K$, which would get a new contribution proportional to the imaginary part of $V^*_{td}V_{ts}$ (for real $[\check{\cC}_{dH}]_{22}$).
We use this connection, alongside the formulae and inputs in App.~\ref{app:meson_mixing}, to place the current indirect constraints:
\begin{equation}
    \kappa_s \lesssim 80~(\Lambda=\text{1\,TeV}), ~~~ \kappa_s \lesssim 50~(\Lambda=\text{3\,TeV}), ~~~ \kappa_s \lesssim 30~(\Lambda=\text{10\,TeV}),
\end{equation}
at 95\% CL. These are weaker than current LHC bounds on $\kappa_s$ for scales below $\sim$10\,TeV, but could place an upper bound on energy scale of new physics should a deviation be observed in $\kappa_s$ with LHC sensitivity (assuming no fine-tuning between different contributions to $\varepsilon_K$).

The SM uncertainty on $\varepsilon_K$ is dominated by the parametric error on the CKM element $|V_{cb}|$. Improvements in the determination of this element at FCC-ee \cite{Marzocca:2024mkc, Liang:2024hox, deBlas:2025gyz} lead us to expect the indirect sensitivity:
\begin{equation}
    \kappa_s \lesssim 30~(\Lambda=\text{1~TeV}), ~~~ \kappa_s \lesssim 20~(\Lambda=\text{3~TeV}), ~~~ \kappa_s \lesssim 10~(\Lambda=\text{10~TeV}).
\end{equation}
at 95\% CL. While significantly weaker than direct sensitivity to $\kappa_s$ at FCC-ee, these bounds could be relevant if a deviation is observed in direct measurements of $\kappa_s$.

In the down sector, indirect constraints from all other flavour observables are significantly weaker than direct bounds. In the up sector, the leading radiative misalignment effect is proportional to $m_b^2/v^2$, so constraints from generated off-diagonal elements are also very weak. We conclude that a reasonable degree of diagonal flavour alignment of Yukawa operators is a necessary condition on any new physics modifying Yukawa couplings at current and future experimental sensitivity. However, should this alignment be achieved in the UV, it is sufficiently radiatively stable that measurable deviations in $\kappa_q$ can occur without conflicts with flavour constraints. 

\subsection{Generating Yukawa modifications radiatively}

If a deviation in $\kappa_q$ is measured, it could also be due to new physics which generates $\hat{\cC}_{uH}$ or $\check{\cC}_{dH}$ radiatively. The one-loop RGE of $\hat{\cC}_{uH}$ is given by (neglecting any SM Yukawa couplings smaller than $y_t$)~\cite{Jenkins:2013wua}
\begin{equation}
\label{eq:rgecuh}
    [\dot{\hat{\cC}}_{uH}]_{ii}\supset\frac{8}{(16\pi^2)}(y_t^3 -\lambda y_t)\left([\hat{\cC}_{qu}^{(1)}]_{i33i} + \frac43[\hat{\cC}_{qu}^{(8)}]_{i33i}\right),
\end{equation}
where $\dot{\cC} \equiv \frac{\text{d}}{\text{d}\ln \mu} \cC$ and we have omitted self renormalisation and terms involving the electromagnetic dipole coefficients $\cC_{uB}$ and $\cC_{uW}$, which cannot be generated at tree level in a weakly-coupled UV completion~\cite{Einhorn:2013kja,Craig:2019wmo}. Potentially significant radiative contributions to the charm and up Yukawa couplings can thus be driven by the following coefficients of four-quark operators:
\begin{equation}
    \kappa_c: [\hat{\cC}_{qu}^{(1)}]_{2332}, [\hat{\cC}_{qu}^{(8)}]_{2332} ~~~~~ \kappa_u: [\hat{\cC}_{qu}^{(1)}]_{1331}, [\hat{\cC}_{qu}^{(8)}]_{1331}. \label{eq:up_type}
\end{equation}
Numerically solving the RGEs using the \texttt{DsixTools} package \cite{Fuentes-Martin:2020zaz}, we find for a UV running scale $\Lambda=3$~TeV
\begin{align}
\kappa_u &= 1 + 560 \,\left([\hat{\cC}_{qu}^{(1)}]_{1331} + 1.3[\hat{\cC}_{qu}^{(8)}]_{1331}\right),\\
\kappa_c &= 1 + 0.97 \left([\hat{\cC}_{qu}^{(1)}]_{2332} + 1.3[\hat{\cC}_{qu}^{(8)}]_{2332}\right), \label{eq:kappa_c}\\
\kappa_t &= 1 + 5.5\times10^{-3} \left([\hat{\cC}_{qu}^{(1)}]_{3333} + 1.4[\hat{\cC}_{qu}^{(8)}]_{3333}\right).
\end{align}
where the Wilson coefficients have units of TeV$^{-2}$ and we have omitted additional coefficients entering into $\kappa_t$. Comparing to Tab.~\ref{tab:kappaprojections}, we see that TeV-scale (or heavier) new physics generating $\hat{\cC}_{qu}^{(1,8)}$ could induce observable deviations in $\kappa_{u,c}$ at HL-LHC or FCC-ee. Deviations in $\kappa_{t}$ from TeV-scale new physics generating these coefficients would not be large enough to be detected, even at FCC-ee.

Analogous analysis of the RGE for the down-type Yukawa coefficient $\check{\cC}_{dH}$~\cite{Jenkins:2013wua}
\begin{equation}
    [\dot{\check{\cC}}_{dH}]_{ii}\supset-\frac{2V_{tb}}{(16\pi^2)}(y_t^3-\lambda y_t)\left(6[\check{\cC}_{quqd}^{(1)}]_{33ii}+[\check{\cC}_{quqd}^{(1)}]_{i33i} + \frac43[\check{\cC}_{quqd}^{(8)}]_{i33i}\right),
\end{equation}
motivates the study of loop-level strange and down Yukawa coupling modifications generated from:
\begin{equation}
    \kappa_s: [\check{\cC}_{quqd}^{(1)}]_{3322},[\check{\cC}_{quqd}^{(1)}]_{2332}, [\check{\cC}_{quqd}^{(8)}]_{2332} ~~~ \kappa_d: [\check{\cC}_{quqd}^{(1)}]_{3311},[\check{\cC}_{quqd}^{(1)}]_{1331}, [\check{\cC}_{quqd}^{(8)}]_{1331}.
    \label{eq:down_type}
\end{equation}
Numerically running from $\Lambda=3$~TeV, we find the Yukawa enhancements
\begin{align}
\kappa_d &= 1 - 400[\check{\cC}_{quqd}^{(1)}]_{3311} - 93 [\check{\cC}_{quqd}^{(1)}]_{1331} - 81 [\check{\cC}_{quqd}^{(8)}]_{1331},\\
\kappa_s &= 1 -19[\check{\cC}_{quqd}^{(1)}]_{3322} - 4.6 [\check{\cC}_{quqd}^{(1)}]_{2332} -4.1 [\check{\cC}_{quqd}^{(8)}]_{2332},\\
\kappa_b &= 1 - 0.54 [\check{\cC}_{quqd}^{(1)}]_{3333} -0.090[\check{\cC}_{quqd}^{(8)}]_{3333},
\end{align}
where again the Wilson coefficients have units of TeV$^{-2}$. In contrast with modifications of up-type Yukawa couplings, observable modifications of the third-generation Yukawa coupling $\kappa_b$ can be induced for TeV-scale new physics generating pure third generation $\check{\cC}_{quqd}^{(1/8)}$. Modifications in the second and first generation may also be accessible at HL-LHC or FCC-ee. 

In the following subsections, we investigate the constraints on these operators from flavour and collider physics, to understand whether they offer a viable mechanism of generating observable Yukawa deviations in future.

\subsection{Constraints on operators generating up-type Yukawa modifications}
We start by calculating constraints on the individual Wilson coefficients $[\hat{\cC}_{qu}^{(1,8)}]_{i33i}$ which can radiatively generate $\kappa_c$ and $\kappa_u$. Then, in Sec.~\ref{sec:general_flavour}, we investigate the possibility of a more general flavour structure for $\hat{\cC}_{qu}^{(1,8)}$, with larger entries in the third generation.

\subsubsection{Top quark pair production}
\label{sec:ttbar}

\begin{figure}
    \centering
    \includegraphics[width=1\linewidth]{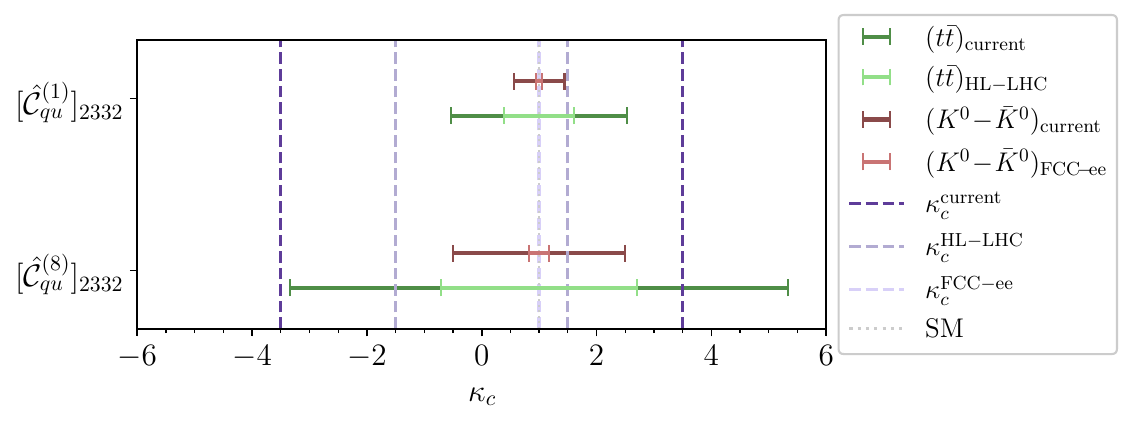}
    \caption{95\% CL allowed regions for $\kappa_c$ deviations induced by the four-quark operator coefficients $\cC_{qu}^{(1,8)}$. The Wilson coefficients are defined at the scale $\Lambda=3$ TeV. Green bars indicate values of $\kappa_c$ allowed by current and projected $t\bar{t}$ production ($\text{d}\sigma/\text{d}m_{t \bar t}$) constraints, assuming only the corresponding Wilson coefficient is non-zero at $\Lambda$. Orange and brick red bars indicate values of $\kappa_c$ allowed by current and projected kaon mixing ($\epsilon_K$) constraints, again assuming only the corresponding Wilson coefficient is non-zero. Vertical dashed lines represent current and projected direct collider constraints on $\kappa_c$ (see Tab.~\ref{tab:kappaprojections}).}
    \label{fig:kappa_c_limits}
\end{figure}

\begin{figure}
    \centering
    \includegraphics[width=1\linewidth]{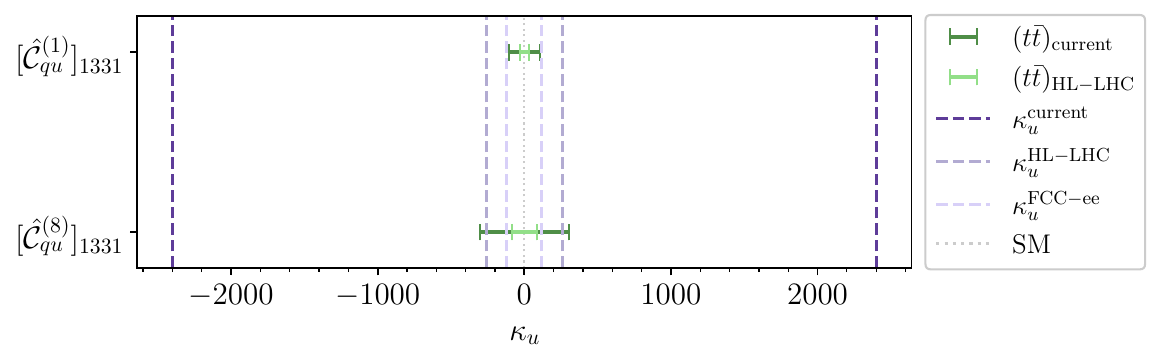}
    \caption{As in Fig.~\ref{fig:kappa_c_limits} but for $\kappa_u$.}
    \label{fig:kappa_u_limits}
\end{figure}

All coefficients in Eq.~\eqref{eq:up_type} contribute to $p p \to t\bar{t}$ at tree level, making LHC $t\bar{t}$ measurements a collider handle on induced $\kappa_c$ or $\kappa_u$ deviations. Their flavour and chirality structure means there is no interference with SM $t\bar{t}$ amplitudes, so the leading SMEFT contribution is a dimension-six squared effect. 

To estimate current bounds from differential LHC $t\bar{t}$ cross section measurements, we calculate the predicted the $t\bar{t}$ cross section at quadratic order in the SMEFT coefficients using \texttt{MadGraph5\_aMC@NLO} \cite{Alwall:2014hca} with the \texttt{SMEFTsim} \cite{Brivio:2017btx, Brivio:2020onw} package. We also provide the leading order parton-level amplitudes and cross sections in Appendix~\ref{app:toppair}.
We work in bins of top pair invariant mass to match those of the CMS analysis \cite{CMS:2021vhb}, combining our SMEFT predictions with the SM prediction used there \cite{Frixione:2007nw, Nason:2004rx, Frixione:2007vw, Alioli:2010xd, Sjostrand:2007gs, Czakon:2011xx}. We compare to the measured CMS $t\bar{t}$ invariant mass distribution ($\text{d}\sigma/\text{d}m_{t \bar t}$) \cite{CMS:2021vhb} and construct a $\chi^2$ function using the provided correlation matrix to derive 95\% confidence intervals on each coefficient individually. 
To determine HL-LHC sensitivity with the full $3~\text{ab}^{-1}$ dataset we use the results of \cite{Durieux:2022cvf}.

These projections account for the greater reach of the full HL-LHC programme through the inclusion of four extra high $m_{t \bar t}$ bins. 
The current and projected HL-LHC constraints from $t\bar t$ production are shown in shades of green in Figs.~\ref{fig:kappa_c_limits} and \ref{fig:kappa_u_limits}, in terms of the allowed size of $\kappa_{c,u}$ that can be generated from the corresponding four-quark Wilson coefficient. Vertical dashed lines in different shades of purple show sensitivities on $\kappa_c$ and $\kappa_u$ from current and future measurements at the LHC and FCC-ee. In Fig.~\ref{fig:kappa_c_limits}, current $t\bar t$ constraints still allow sizeable enhancements in $\kappa_c$ that could be observable at HL-LHC, particularly in the case of a contribution from $[\hat{\cC}_{qu}^{(8)}]_{2332}$.

Turning to $\kappa_u$ enhancements, shown in Fig.~\ref{fig:kappa_u_limits}, the allowed parameter space is much more tightly constrained by $t\bar{t}$ measurements due to the parton distribution function (PDF) enhancement relative to the charm case. 
At HL-LHC, measurements of $t\bar{t}$ production will test these four-quark coefficients better than $\kappa_u$, and if no deviation is seen in $t\bar{t}$ measurements at HL-LHC, then any observed deviation in $\kappa_u$ at FCC-ee cannot be (entirely) due to these four-quark interactions.

\subsubsection{Flavour constraints: kaon mixing}
\label{sec:kaonmix}
Induced enhancements of $\kappa_c$ receive key constraints from $\varepsilon_{K}$. At leading order, this is driven by the RGE terms~\cite{Jenkins:2013wua}
\begin{align}
    [\dot{\check{\cC}}^{(1)}_{qq}]_{1212}&\supset -\frac{V_{cd}^{*}V_{ts} V_{cs}V_{td}^{*} y_t y_c}{(16\pi^2)} \left([\hat{\cC}^{(1)}_{qu}]_{2332}+\frac{1}{12}[\hat{\cC}^{(8)}_{qu}]_{2332}\right),\\
    [\dot{\check{\cC}}^{(3)}_{qq}]_{1212}&\supset -\frac{V_{cd}^{*}V_{ts} V_{cs}V_{td}^{*} y_t y_c}{4(16\pi^2)}\, [\hat{\cC}^{(8)}_{qu}]_{2332}.
\end{align}
Since the CKM prefactor has an imaginary part, and $\mathrm{Im}[\check{\cC}^{(1)}_{qq}]_{1212}$ contributes to $\varepsilon_{K}$ at tree level, this observable puts constraints on $[\hat{\cC}^{(1,8)}_{qu}]_{2332}$ generated above the EW scale. We re-sum the relevant RGEs using using both \texttt{wilson} \cite{Aebischer:2018bkb} and \texttt{DsixTools} \cite{Fuentes-Martin:2020zaz}, and use the formulae and inputs given in App.~\ref{app:meson_mixing} to constrain the coefficients $[\hat{\cC}^{(1,8)}_{qu}]_{2332}$. There are no analogous constraints on $[\hat{\cC}^{(1,8)}_{qu}]_{1331}$ due to the additional Yukawa suppression.

We show these constraints, in terms of the maximum $\kappa_c$ which can be induced by the corresponding Wilson coefficient, in brown in Fig.~\ref{fig:kappa_c_limits}. Projected FCC-ee sensitivities are shown in orange, where the improvement comes from an improved determination of $V_{cb}$ from $W$ decays \cite{Marzocca:2024mkc, Liang:2024hox}. For $\hat{\cC}_{qu}^{(1)}$, we find that $\kappa_c$ enhancements at the level of current and HL-LHC sensitivity are already forbidden by current $\varepsilon_K$ measurements. At FCC-ee, a direct measurement of $\kappa_c$ is the best probe of the parameter space, although improved precision on $\varepsilon_K$ will lead to competitive indirect constraints.

For $\hat{\cC}_{qu}^{(8)}$, a smaller contribution to both kaon mixing and $t \bar t$ leaves room for loop-induced $\kappa_c$ enhancements that may be observable throughout the LHC and HL-LHC programmes. At FCC-ee, a direct measurement of $\kappa_c$ is the strongest probe of this coefficient, surpassing any correlated constraints. We conclude that at both HL-LHC and FCC-ee, new physics contributing to $[\hat{\cC}_{qu}^{(8)}]_{2332}$ could first make itself known as a deviation in $\kappa_c$.

\subsubsection{Third-generation-philic flavour scenario}
\label{sec:general_flavour}
To understand the full reach of HL-LHC and FCC-ee $\kappa_{u,c}$ sensitivity, we now investigate how generic the flavour structure of $\hat{\cC}_{qu}^{(1)}$ or $\hat{\cC}_{qu}^{(8)}$ can be without inducing measurable deviations in complementary observables. Motivated by both the flavour structure of the SM Yukawa matrices and the generally stronger experimental bounds on new physics involving light quarks, we consider flavour structures in which new physics couples preferentially to the third generation. Similar scenarios have been extensively studied in the SMEFT, e.g.\ in the context of an approximate $U(2)^n$ flavour symmetry in the first two generations of quarks and/or leptons (e.g.~\cite{Faroughy:2020ina, Allwicher:2023shc, Allwicher:2025bub}). However in the case we are investigating this symmetry assumption cannot be applied without eliminating contributions to $\kappa_{c,u}$, since a $[\Op_{qu}^{(1,8)}]_{i33i}$ operator (with $i=1,2$) is not invariant under $U(2)_q\times U(2)_u$.

\begin{figure}
    \centering
    \begin{subfigure}[b]{0.40\textwidth}
        \centering
        \includegraphics[width=\textwidth]{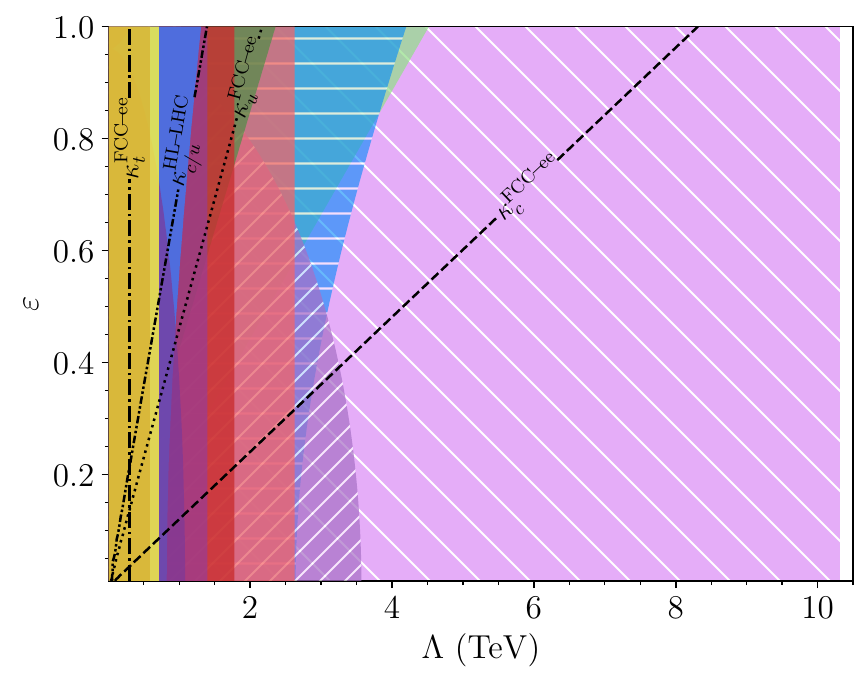}
        \caption{$\mathbf{\cC_{qu}^{(1)}}$}
        \label{fig:full_flav_cqu1}
    \end{subfigure}
    \hfill
    \begin{subfigure}[b]{0.545\textwidth}
        \centering
        \includegraphics[width=\textwidth]{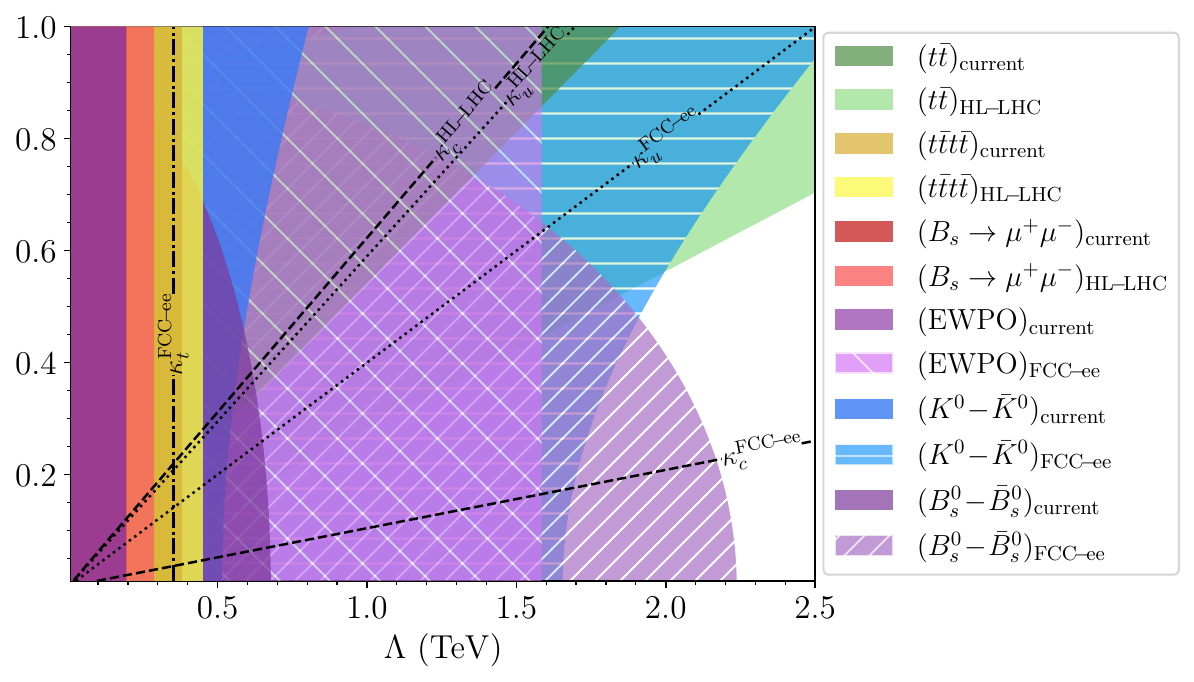}
        \caption{$\mathbf{\cC_{qu}^{(8)}}$}
        \label{fig:full_flav_cqu8}
    \end{subfigure}
    \caption{Parameter space for up-type Yukawa enhancements assuming the flavour structure of Eq.~\eqref{eq:diagflavassump} and only $\cC_{qu}^{(1)}$ (a) or $\cC_{qu}^{(8)}$ (b) is non-zero at the UV scale. Shaded regions are excluded at 95\% confidence by the corresponding current or projected measurement. Dashed and dotted lines indicate points in parameter space where Yukawa enhancements are equal to the experimental constraints in Tab.~\ref{tab:kappaprojections}. Note in (b) the current $B_s\to \mu^+\mu^-$ region lies almost exactly underneath the current EWPO region.}
    \label{fig:full_flavour_structure}
\end{figure}

Instead we start from a general ansatz where at the UV matching scale each $[\hat{\cC}_{qu}^{(1,8)}]_{ijkl}$ is related to the all-third generation coefficient $[\hat{\cC}_{qu}^{(1,8)}]_{3333}$ by parametric suppression of $\varepsilon<1$ for each non-third-generation flavour index, i.e.
\begin{equation}
    [\hat{\cC}_{qu}^{(1,8)}]_{ijkl}=\varepsilon^{n_q}[\hat{\cC}_{qu}^{(1,8)}]_{3333},
    \label{eq:flav_struc}
\end{equation}
where $n_q$ is the number of first and second generation quark fields in the operator $[\hat{\Op}_{qu}^{(1,8)}]_{ijkl}$. This flavour structure is similar to those found in models of partial compositeness \cite{Kaplan:1991dc}, for example. In Fig.~\ref{fig:full_flavour_structure}, we show the sensitivity reach of current and projected $\kappa_q$ limits in the plane of $\varepsilon$ and $\Lambda$, where $\Lambda=[\hat{\cC}_{qu}^{(1,8)}]_{3333}^{-1/2}$. A measurement is sensitive to all regions above and to the left of the corresponding line, at 95\% CL.\footnote{The $\kappa_q$ constraints and projections should be taken as indicative here, since some analyses assume SM-like values for other $\kappa$s, which does not hold in this plane since $\kappa_u$, $\kappa_c$ and $\kappa_t$ all vary.} Note that here we have made the assumption of a common suppression $\varepsilon$ for both first- and second-generation quarks, such that both $\kappa_u$ and $\kappa_c$ appear on this plane.

While the flavour ansatz \eqref{eq:flav_struc} allows us to explore a broad class of UV flavour structures, in its most general form it simultaneously introduces a set of strong and unavoidable constraints.
In particular, flavour components which are not consistent with a $U(1)_{t}\times U(1)_{c}\times U(1)_{u}$ vectorial flavour symmetry receive very strong constraints from meson mixing. For instance, $[\hat{\cC}_{qu}^{(1)}]_{1233}$ contributes to $D$-meson mixing at one-loop; current constraints therefore already place an effective bound on $\kappa_c$ sixty times stronger than projected direct sensitivity at FCC-ee \cite{Silvestrini:2018dos}. We therefore impose this further vectorial symmetry on the coefficients at the matching scale, as a device to estimate the minimal off-diagonal effects.\footnote{Given that this is not an exact symmetry in the SM, this assumption is not radiatively stable and receives corrections proportional to products of CKM elements. We include these radiative corrections in the calculation of the bounds, running from the NP scale $\Lambda$.}
This means that we assume that the only non-zero coefficients (at the UV scale) are:
\begin{equation}
\label{eq:diagflavassump}
\mathbf{U(1)_{t}\times U(1)_{c}\times U(1)_{u}}
\quad
\!
~~
\begin{aligned}
    [\hat{\cC}_{qu}^{(1,8)}]_{i33i}=[\hat{\cC}_{qu}^{(1,8)}]_{ii33}=[\hat{\cC}_{qu}^{(1,8)}]_{33ii}=\varepsilon^2[\hat{\cC}_{qu}^{(1,8)}]_{3333} \\
    [\hat{\cC}_{qu}^{(1,8)}]_{iiii}=[\hat{\cC}_{qu}^{(1,8)}]_{iijj}=[\hat{\cC}_{qu}^{(1,8)}]_{ijji}=\varepsilon^4[\hat{\cC}_{qu}^{(1,8)}]_{3333},
\end{aligned}
\end{equation}
where $i,j=1,2$. This is the flavour assumption that goes into the constraints in Fig.~\ref{fig:full_flavour_structure}, which are described in turn below. 

The third generation-philic nature of the assumptions Eq.~\eqref{eq:diagflavassump} gives importance to the phenomenology of $[\hat{\cC}_{qu}^{(1)}]_{3333}$ and $[\hat{\cC}_{qu}^{(8)}]_{3333}$.\footnote{As a by-product of our analysis here, we summarise the various constraints on fully third generation operator coefficients in App.~\ref{app:thirdgen}.} At tree-level, these coefficients enter four top production $(t \bar{t} t \bar{t})$ which we calculate using \texttt{MadGraph5\_aMC@NLO} \cite{Alwall:2014hca} with \texttt{SMEFTsim} \cite{Brivio:2017btx, Brivio:2020onw}. We combine these with the SM prediction of \cite{vanBeekveld:2022hty} and compare to the ATLAS and CMS inclusive cross section measurements taken at at $\sqrt{s}=13$~TeV \cite{ATLAS:2023ajo, CMS:2023ftu}. These bounds are shown in yellow (current) and mustard (HL-LHC projected \cite{ATLAS:2018kci}) in Fig.~\ref{fig:full_flavour_structure}.

At one loop, $[\hat{\cC}_{qu}^{(1)}]_{3333}$ and $[\hat{\cC}_{qu}^{(8)}]_{3333}$ also modify the branching ratio $\mathcal{B}(B_s \to \mu^+ \mu^-)$. We calculate constraints from this process using \texttt{wilson} \cite{Aebischer:2018bkb} and \texttt{DsixTools} \cite{Fuentes-Martin:2020zaz}, alongside the SM predictions and experimental measurements from \cite{Czaja:2024the, ATLAS:2025lrr, ParticleDataGroup:2026aaa}. We summarise the relevant formulae and inputs in App.~\ref{app:bsmumu_formulae}. These constraints are shown in brick red (current) and salmon (HL-LHC projected) in Fig.~\ref{fig:full_flavour_structure}. 

There are also constraints from the $\Delta m_{B_s}$ observable from $B_s$ mixing, which are strongest when $\varepsilon\to 0$, since for non-zero values of $\varepsilon$ there is a partial GIM cancellation among the leading vectorial pieces $[\hat{\cC}_{qu}^{(1,8)}]_{ii33}$, which becomes exact when $\varepsilon=1$. We calculate these constraints using the formulae and method outlined in App.~\ref{app:meson_mixing}. The $\Delta m_{B_s}$ constraints are shown in dark purple (current~\cite{utfit}) and lighter purple (FCC-ee, where the improvement is mostly from an improved determination of $V_{cb}$~\cite{Charles:2020dfl,deBlas:2025gyz}).

Finally, at one- and two-loop, these fully third generation coefficients also contribute to electroweak precision observables. At one-loop, $[\hat{\cC}_{qu}^{(1)}]_{3333}$ contributes to $Z\to \bar b_L b_L$, while at two loop it gives contributions to the $T$ parameter. We evaluate these contributions using recent two-loop calculations of the observables in terms of SMEFT Wilson coefficients~\cite{Haisch:2024wnw},
and find constraints by fitting to the full set of $Z$ pole observables from LEP and SLD~\cite{ALEPH:2005ab,Janot:2019oyi}, as well as the latest average of the $W$ mass~\cite{LHC-TeVMWWorkingGroup:2023zkn}. The resulting 95\% CL constraints are shown in purple in Fig.~\ref{fig:full_flavour_structure}. \emph{It can be seen that, under the flavour assumption of Eq.~\eqref{eq:diagflavassump} with $\epsilon<1$, electroweak precision constraints are enough to rule out deviations in $\kappa_c$ and $\kappa_u$ from $\hat{\cC}_{qu}^{(1)}$ at HL-LHC}. At FCC-ee, the huge number of $Z$ bosons produced will enormously improve on the sensitivity. The hatched lilac regions show projected sensitivity taking experimental projections from Ref.~\cite{Selvaggi:2025kmd} and assuming the `aggressive' scenario for theoretical errors described in Ref.~\cite{deBlas:2025gyz}. This will fully test the region of FCC-ee sensitivity to $\hat{\cC}_{qu}^{(1)}$ in $\kappa_c$, assuming the flavour assumption of Eq.~\eqref{eq:diagflavassump}. Turning to enhancements induced by $\hat{\cC}_{qu}^{(8)}$, its colour structure introduces an additional loop suppression in contributions to EWPOs, rendering those constraints comparatively weak. FCC-ee $\kappa_c$ measurements emerge as the most powerful future probe of this coefficient.

At $\mathcal{O}(\varepsilon^2)$, there are four allowed entries involving vector currents of both top quarks and light quarks: $[\hat{\mathcal{C}}_{qu}^{(1,8)}]_{33ii}$ and $[\hat{\mathcal{C}}_{qu}^{(1,8)}]_{ii33}$, as well as the two previously discussed entries involving scalar currents, $[\hat{\mathcal{C}}_{qu}^{(1,8)}]_{i33i}$. The colour-octet vector-current operators contribute to $t \bar t$ production at tree level where, unlike the scalar current case, there is interference between SM and SMEFT amplitudes (see App.~\ref{app:toppair}). Existing and projected (HL-LHC) $t\bar{t}$ constraints are shown in shades of green in Fig.~\ref{fig:full_flavour_structure}, calculated as described in Sec.~\ref{sec:ttbar}. Overall it can be seen that for either coefficient, Yukawa enhancements at HL-LHC sensitivity are already fully excluded by current $t \bar t$ measurements, where the constraint is driven by flavour components involving first generation quarks. 

Coefficients of operators with a vector current of right-handed tops, i.e. $[\hat{\mathcal{C}}_{qu}^{(1,8)}]_{ii33}$ contribute to $K^0$--$\bar{K}^0$ mixing ($\varepsilon_K$) at one loop. However, with equal coefficients for $i=1$ and $i=2$, these operators inherit a $U(2)$ flavour symmetry in the first two generations which leads to substantial cancellations, suppressing these contributions. The leading $K^0$--$\bar{K}^0$ constraints are therefore from the scalar current operators $[\hat{\mathcal{C}}_{qu}^{(1,8)}]_{2332}$, as described in Sec.~\ref{sec:kaonmix} above. These are shown in shades of blue in Fig.~\ref{fig:full_flavour_structure}, where the improvement at FCC-ee is from increased precision on $V_{cb}$, which will reduce theoretical errors.

\begin{figure}
    \centering
    \begin{subfigure}[b]{0.4\textwidth}
        \centering
        \includegraphics[width=\textwidth]{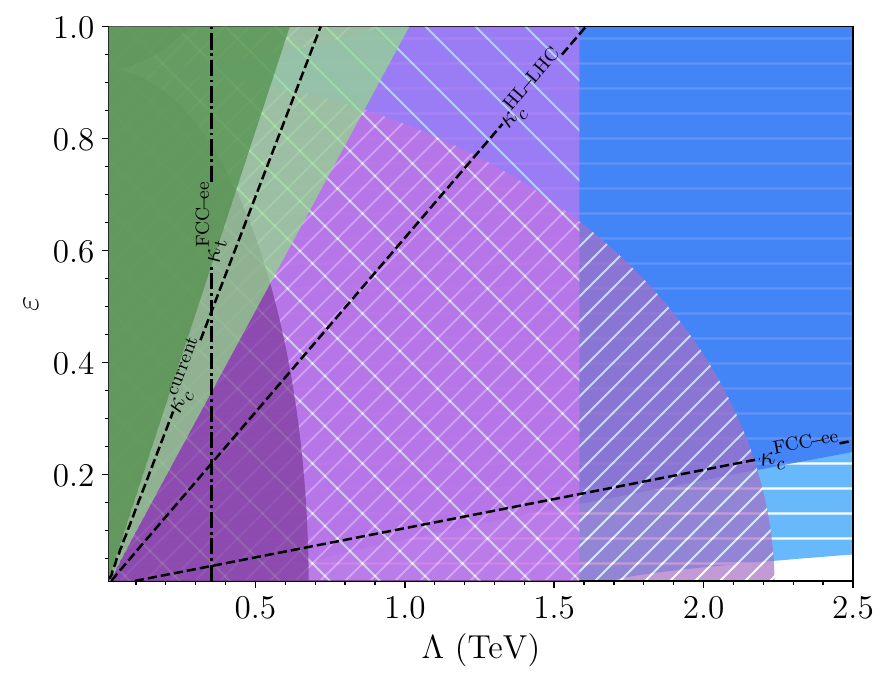}
        \caption{}
        \label{fig:full_charm_cqu8}
    \end{subfigure}
    \hfill
    \begin{subfigure}[b]{0.521\textwidth}
        \centering
        \includegraphics[width=\textwidth]{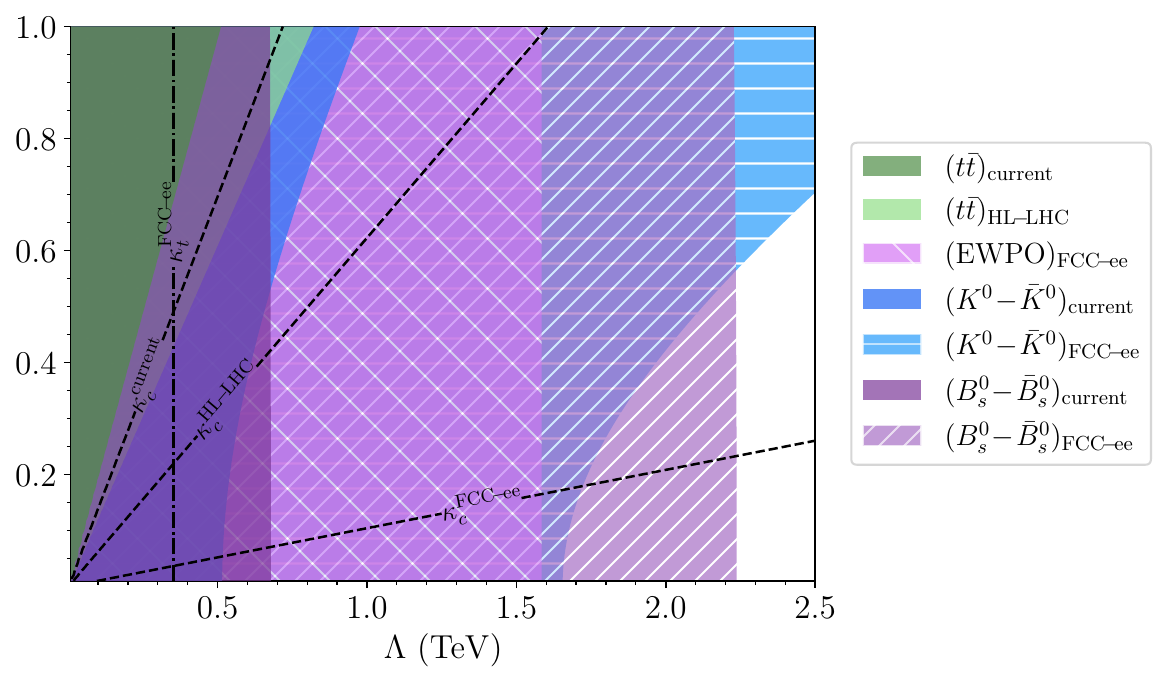}
        \caption{}
        \label{fig:partial_charm_cqu8}
    \end{subfigure}
    \caption{Parameter space assuming only $\cC_{qu}^{(8)}$ at the UV scale, with the flavour structure of Eq.~\eqref{eq:diagflavassumpnofirstgen} (a) or Eq.~\eqref{eq:scalarsonly} (b). To simplify the plots, subleading constraints on the third generation components are not shown here, but are identical to those shown in Fig.~\ref{fig:full_flav_cqu8}.}
    \label{fig:cqu8restricted}
\end{figure}

The $t\bar t$ constraints are unavoidable when enhancing $\kappa_u$, however, in analogy with the flavour structure of the SM, it is not unnatural to expect coefficients with first generation flavour indices to be further suppressed relative to coefficients with second generation indices. In Fig.~\ref{fig:full_charm_cqu8}, we show the parameter space in the extremal case where all first-generation flavour indices are forbidden. Specifically, we have 
\begin{equation}
\label{eq:diagflavassumpnofirstgen}
\begin{aligned}
\mathbf{U(1)_{t}\times U(1)_{c}~}\\
\mathrm{\textbf{(no first gen)}}
\end{aligned}
\quad
\!
~~~~~
\begin{aligned}
    [\hat{\cC}_{qu}^{(1,8)}]_{2332}=[\hat{\cC}_{qu}^{(1,8)}]_{2233}=[\hat{\cC}_{qu}^{(1,8)}]_{3322}=\varepsilon^2[\hat{\cC}_{qu}^{(1,8)}]_{3333} \\ [\hat{\cC}_{qu}^{(1,8)}]_{2222}=\varepsilon^4[\hat{\cC}_{qu}^{(1,8)}]_{3333},
\end{aligned}
\end{equation}
with all other coefficients zero.
Under this assumption, the $U(2)$ flavour symmetry among the vector current $\mathcal{O}(\varepsilon^2)$ coefficients is broken, rendering $K^0$--$\bar{K}^0$ mixing constraints significant. Current measurements of $\varepsilon_{K}$ already exclude $\kappa_c$ deviations at the level of FCC-ee sensitivity, while anticipated improvements in the $\varepsilon_{K}$ determination will strengthen these bounds further. Similar flavour assumptions with varying suppression between first and second generation indices, will interpolate between these two scenarios (in Fig.~\ref{fig:full_charm_cqu8} and Fig.~\ref{fig:full_flavour_structure} (b)).

Instead, if UV physics only populates operators involving \emph{scalar currents} of tops and charms, such strong constraints from kaon mixing are avoided. In this case, the only non-zero coefficients are
\begin{equation}
\label{eq:scalarsonly}
\begin{aligned}
\mathrm{\textbf{Scalar currents only}}\\
\mathrm{\textbf{(No first gen)}}
\end{aligned}
\quad
\!
~~~~~~~
\begin{aligned}
[\hat{\cC}_{qu}^{(1,8)}]_{2332}=\varepsilon^2[\hat{\cC}_{qu}^{(1,8)}]_{3333}, \\ [\hat{\cC}_{qu}^{(1,8)}]_{2222}=\varepsilon^4[\hat{\cC}_{qu}^{(1,8)}]_{3333}.
\end{aligned}
\end{equation}
We show this scenario in Fig.~\ref{fig:partial_charm_cqu8}, where measurements of $\kappa_c$ at HL-LHC and FCC-ee will test new parameter space 

We conclude that third-generation-philic new physics with a hierarchy $\varepsilon \gtrsim 0.2$, could first make itself known through measurements of $\kappa_c$ at FCC-ee. This is conditional on generating $\hat{\cC}_{qu}^{(8)}$ with a generational vectorial symmetry. 

For deviations to be induced at HL-LHC sensitivity, without violating $K^0$-$\bar{K}^0$ or $t \bar{t}$ constraints, a third-generation-philic UV must only populate the top/light-quark scalar current entries of $\hat{\cC}_{qu}^{(8)}$, with some suppression of first generation couplings with respect to the second. In this case, $\kappa_c$ modifications at HL-LHC sensitivity are achievable for moderate inter-generational hierarchies $\varepsilon \gtrsim 0.4$.

The required hierarchies for enhancing $\kappa_c$ to match future experimental precision, without incurring correlated constraints, are still far from well known paradigms such as Minimal Flavour Violation (MFV). Remaining agnostic about the nature of new physics, we nevertheless identify a target SMEFT region, albeit with several flavour restrictions, that motivates further experimental focus on $\kappa_c$.

\subsection{Constraints on operators generating down-type Yukawa modifications}
\label{sec:down_type_eft}
While the operators in Eq.~\eqref{eq:down_type} induce one-loop modifications of down-type Yukawa couplings, they also modify radiative $b\to s(d) \gamma$ processes at loop level. Measurements of the branching ratios $\mathcal{B}(B\to X_{s}\gamma)$ and $\mathcal{B}(B\to X_{d}\gamma)$ therefore provide strong constraints on the relevant coefficients. As was the case for up-type enhancements, the four-quark operator structures also contribute to $t \bar t$ production, providing complementary collider constraints. There are no meaningful constraints from meson mixing on these operators, since the contributions are suppressed by $y_b$.

We calculate $\mathcal{B}(B\to X_{s}\gamma)$ for a UV scale $\Lambda=3$~TeV, using \texttt{wilson} for one-loop RGE evolution and tree-level matching to the relevant low-energy coefficients. We use the formulae and inputs from \cite{Bruggisser:2021duo, Bause:2022rrs, Misiak:2020vlo} (summarised in App.~\ref{app:bXdgamma_formulae}) to derive constraints from the current experimental average for $\mathcal{B}(B\to X_{s}\gamma)$ \cite{HFLAV:2024ctg} and the extrapolated value for $\mathcal{B}(B\to X_{d}\gamma)$ \cite{Misiak:2015xwa, BaBar:2010vgu}. If Belle~II reaches 50~ab$^{-1}$ of integrated luminosity, the experimental uncertainties are projected to decrease to 4.7\% \cite{ATLAS:2025lrr} and 14\% \cite{Belle-II:2018jsg} respectively.
Constraints from $t \bar t$ production are derived following the procedure of Sec.~\ref{sec:ttbar}. 

\begin{figure}
    \centering
    \includegraphics[width=1\linewidth]{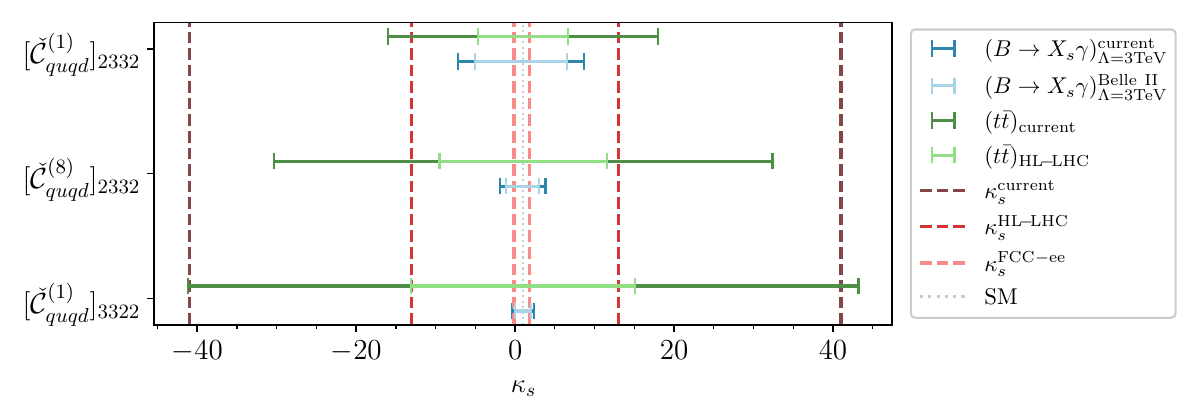}
    \caption{95\% CL allowed regions for $\kappa_s$ deviations induced by the four-quark operator coefficients $\check{\cC}_{quqd}^{(1,8)}$. The Wilson coefficients are defined at the scale $\Lambda=3$ TeV. Green bars indicate values of $\kappa_s$ allowed by current and projected $t\bar{t}$ production ($\text{d}\sigma/\text{d}m_{\bar tt}$) constraints, assuming only the corresponding Wilson coefficient is non-zero at $\Lambda$. Blue bars indicate values of $\kappa_s$ allowed by current and projected $B\to X_s \gamma$ constraints, again assuming only the corresponding Wilson coefficient is non-zero. Vertical dashed lines represent current and projected direct collider constraints on $\kappa_s$ (see Tab.~\ref{tab:kappaprojections}).}
    \label{fig:kappa_s}
\end{figure}

Beginning with $\kappa_s$ enhancements, shown in Fig.~\ref{fig:kappa_s}, we find that current measurements of $\mathcal{B}(B\to X_{s}\gamma)$ place strong constraints, ruling out Yukawa deviations accessible at HL-LHC. Constraints from $t \bar t $ measurements are subdominant, with the exception of $[\check{\cC}_{quqd}^{(1)}]_{2332}$, where HL-LHC projections are competitive with flavour. FCC-ee sensitivity to $\kappa_s$ is sufficient to probe beyond the flavour bounds for both $[\check{\cC}_{quqd}^{(1)}]_{2332}$ and $[\check{\cC}_{quqd}^{(8)}]_{2332}$, leaving a small region of parameter space in which enhancements are viable.

\begin{figure}
    \centering
    \includegraphics[width=1\linewidth]{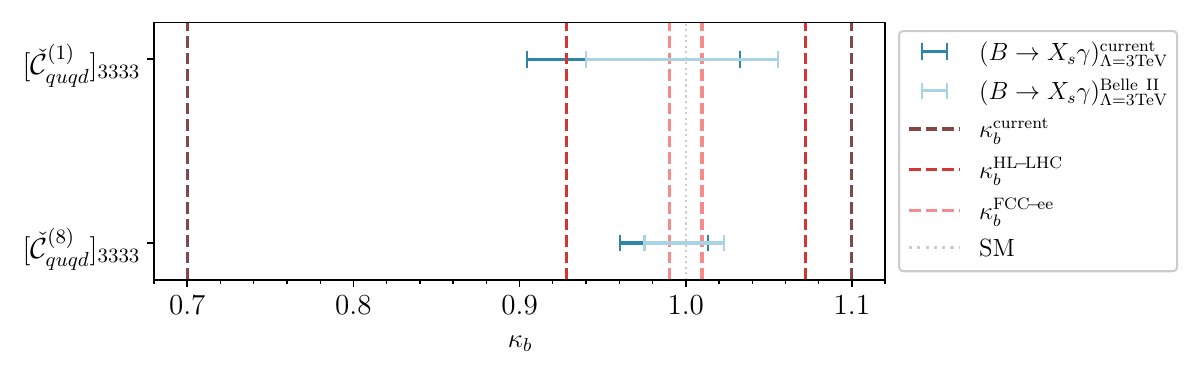}
    \caption{As in Fig.~\ref{fig:kappa_s} but for $\kappa_b$. Measurements of $t\bar t$ production do not provide meaningful constraints here.}
    \label{fig:kappa_b}
\end{figure}

Turning to $\kappa_b$ enhancements, shown in Fig.~\ref{fig:kappa_b}, weaker flavour constraints allow for $\kappa_b$ deviations at HL-LHC sensitivity for $[\check{\cC}_{quqd}^{(1)}]_{3333}$. FCC-ee sensitivity to $\kappa_b$ will provide the best probe of both $[\check{\cC}_{quqd}^{(1)}]_{3333}$ and $[\check{\cC}_{quqd}^{(8)}]_{3333}$. We summarise the flavour bounds on these all-third-generation coefficients in App.~\ref{app:thirdgen}. 

\begin{figure}
    \centering
    \includegraphics[width=1\linewidth]{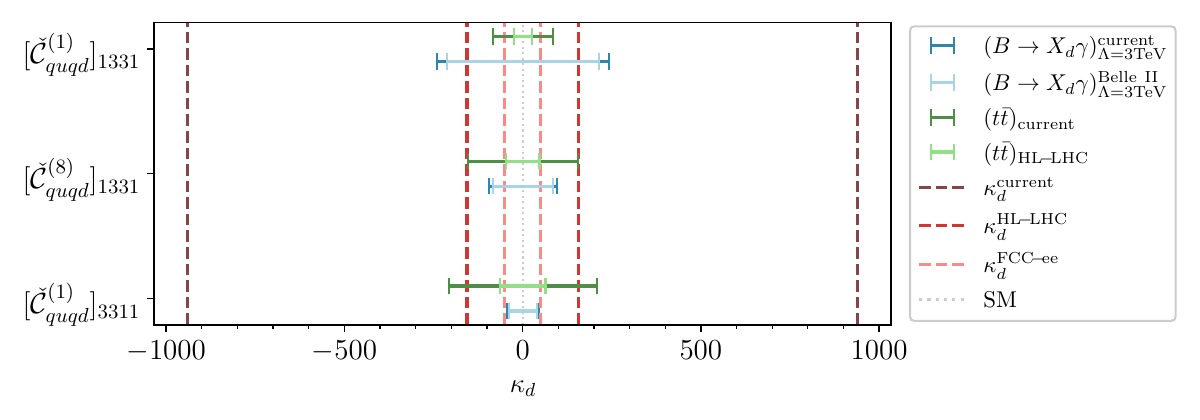}
    \caption{As in Fig.~\ref{fig:kappa_s} but for $\kappa_d$.}
    \label{fig:kappa_d}
\end{figure}

Induced enhancements to $\kappa_d$, shown in Fig.~\ref{fig:kappa_d}, are strongly constrained by measurements of $\mathcal{B}(B\to X_{d}\gamma)$ and $\text{d}\sigma/\text{d}m_{t \bar t}$, where contributions to the latter are enhanced by the relatively large $d$-quark PDF. Taken together, these observables exclude $\kappa_d$ deviations from these operators at HL-LHC. If no anomalous signals are seen in $t\bar{t}$ distributions at HL-LHC, then $\kappa_d$ deviations will also be excluded to FCC-ee sensitivities.

So the most promising flavour scenario for testing four-quark operators in HL-LHC down-type Yukawa measurements is a deviation in $\kappa_b$ driven by the all-third-generation coefficient $[\check{\cC}_{quqd}^{(1)}]_{3333}$. This coefficient is allowed by a $U(2)^3$ flavour symmetry structure on the new physics, under which all third generation quarks are singlets.

\subsection{Summary and discussion}
Summarising the situation in the SMEFT, we find that in many cases four-quark operators are well constrained by flavour and/or top pair measurements, such that they should not be expected to produce observable deviations in $\kappa_q$ at HL-LHC. This is the case for all coefficients that can affect $\kappa_{u}$, $\kappa_{d}$ and $\kappa_{s}$. If a deviation in any of these couplings is observed at HL-LHC, then we can be confident that the (heavy, weakly coupled) BSM explanation is a model which matches onto $\cC_{qH}$ at tree level, such as one of the models in \cite{Giannakopoulou:2024unn, Altmannshofer:2016zrn,Erdelyi:2024sls}. There is nevertheless still room for deviations in all of these via four-quark operators at FCC-ee sensitivity, within current bounds.

For $\kappa_c$ and $\kappa_b$, however, deviations at HL-LHC sensitivity could be due to an underlying four-quark interaction. $\kappa_b$ can be enhanced by the all-third-generation coefficients $[\check{\cC}_{quqd}^{(1,8)}]_{3333}$, while $\kappa_c$ enhancements could be due to $[\hat{\cC}_{qu}^{(1,8)}]_{2332}$. In Sec.~\ref{sec:general_flavour} we explored the possibility of a third-generation-philic flavour structure for $\hat{\cC}_{qu}^{(1,8)}$, which nevertheless still shows up first in measurements of $\kappa_c$. We found that this is not easy to achieve, due to competitive constraints from meson mixing and electroweak precision, but that it is possible if new physics matches mostly onto $\hat{\cC}_{qu}^{(8)}$ and generates no vector currents of second generation quarks.

In the next Section, we explore simple single-particle UV completions for these operators, to determine whether the particular flavour and helicity structures needed are achievable in models of new physics without incurring additional strong constraints.

\section{Simple UV completions for Yukawa modifications}
\label{sec:uv_models}
In the previous Section, we explored the full space of four-quark operators which can generate light-quark Yukawa deviations radiatively, finding $\kappa_c$ and $\kappa_b$ to be particularly sensitive probes. To understand how the viable parameter space can be populated, we now study simplified SM extensions, which match at tree level to the SMEFT, using the results of \cite{deBlas:2017xtg}. We define our new physics Lagrangian identically to this work and adopt their notation for states and couplings. As in the previous section, we assume all new physics couplings are purely real at the UV scale. Each explicit model generates more than one Wilson coefficient thereby introducing additional constraints. Direct searches for the states also provide key probes of the relevant parameter space. 

\subsection{Single-particle completions for $\kappa_c$}
\begin{figure}
    \centering
    \includegraphics[width=0.6\linewidth]{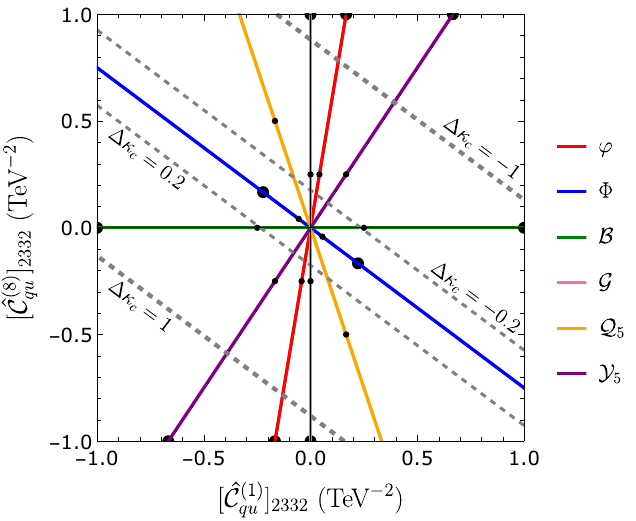}
    \caption{Plot depicting the matching of the six states listed in Tab.~\ref{tab:UVstates} onto the coefficients $[\hat \cC_{qu}^{(1,8)}]_{2332}$. The larger (smaller) black blobs on each line show the points for which the mass of the state is at $1$ TeV ($2$ TeV), assuming couplings of unit magnitude. Contours of $\Delta \kappa_c$ are also shown (with $\Delta \kappa_c \equiv \kappa_c -1$), assuming running from a scale of $\Lambda=2$ TeV.}
    \label{fig:statesplot}
\end{figure}

We gather all the states which match to $[\hat{\cC}_{qu}^{(1,8)}]_{2332}$ at tree level in Tab.~\ref{tab:UVstates}, noting also the other coefficients which are inevitably generated by the relevant couplings. In Fig.~\ref{fig:statesplot}, we show the matching relations for each state in the plane of the relevant coefficients, alongside contours of $\kappa_c$ and benchmark enhancements for unit couplings. 

\begin{table}[]
    \centering
    \begin{tabular}{c|c|c|c|c|c}
    State & Spin & QNs & $[\hat{\cC}_{qu}^{(1)}]_{2332}$ & $[\hat{\cC}_{qu}^{(8)}]_{2332}$ & Other coeffs\\
    \hline
      $\varphi$ & $0$ & $(1,2,-\frac{1}{2})$ & $-\frac{(y_\varphi^u)^*_{33}(y_\varphi^u)_{22}}{6 M_\varphi^2}$ & $-\frac{(y_\varphi^u)^*_{33}(y_\varphi^u)_{22}}{M_\varphi^2}$ & $[\hat{\cC}_{qu}^{(1,8)}]_{3333}$, $[\hat{\cC}_{qu}^{(1,8)}]_{2222}$ \\
       $\Phi$  & $0$ & $(8,2,-\frac{1}{2})$ & $-\frac{2(y_\Phi^{qu})^*_{33}(y_\Phi^{qu})_{22}}{9 M_\Phi^2}$ & $\frac{(y_\Phi^{qu})^*_{33}(y_\Phi^{qu})_{22}}{6 M_\Phi^2}$ & $[\hat{\cC}_{qu}^{(1,8)}]_{3333}$, $[\hat{\cC}_{qu}^{(1,8)}]_{2222}$ \\
       $\mathcal{B}$ & 1 & $(1,1,0)$ & $-\frac{(g_{\mathcal{B}}^{u})_{32}(g_{\mathcal{B}}^{q})_{23}}{M_{\mathcal{B}}^2}$ & 0 & $[\hat{\cC}_{qq}^{(1)}]_{2323}$, $[\hat{\cC}_{uu}]_{3232}$, $[\hat{\cC}_{qu}^{(1)}]_{2323}$\\
       $\mathcal{G}$ & 1 & $(8,1,0)$ & 0 &$-\frac{(g_{\mathcal{G}}^{u})_{32}(g_{\mathcal{G}}^{q})_{23}}{M_{\mathcal{G}}^2}$ & $[\hat{\cC}_{qq}^{(1, 3)}]_{2323}$, $[\hat{\cC}_{uu}]_{3232}$, $[\hat{\cC}_{qu}^{(8)}]_{2323}$\\
       $\mathcal{Q}_5$ & 1 & $(3,2,-\frac{5}{6})$ & $\frac{2(g_{\mathcal{Q}_5}^{uq})^*_{23}(g_{\mathcal{Q}_5}^{uq})_{32}}{3M_{\mathcal{Q}_5}^2}$ & $-\frac{2(g_{\mathcal{Q}_5}^{uq})^*_{23}(g_{\mathcal{Q}_5}^{uq})_{32}}{M_{\mathcal{Q}_5}^2}$ &  $[\hat{\cC}_{qu}^{(1,8)}]_{2233}$, $[\hat{\cC}_{qu}^{(1,8)}]_{3322}$  \\
       $\mathcal{Y}_5$ & 1 & $(\bar{6},2,-\frac{5}{6})$ & $\frac{2(g_{\mathcal{Y}_5})^*_{23}(g_{\mathcal{Y}_5})_{32}} {3M_{\mathcal{Y}_5}^2}$ & $\frac{(g_{\mathcal{Y}_5})^*_{23}(g_{\mathcal{Y}_5})_{32}}{M_{\mathcal{Y}_5}^2}$&  $[\hat{\cC}_{qu}^{(1,8)}]_{2233}$, $[\hat{\cC}_{qu}^{(1,8)}]_{3322}$ \\
    \end{tabular}
    \caption{States generating $[\hat{\cC}_{qu}^{(1,8)}]_{2332}$ at tree level, their quantum numbers (QNs) and their tree-level matching, from Ref.~\cite{deBlas:2017xtg}. The final column gives other coefficients and flavour structures that are inevitably generated by the same couplings which generate $[\hat{\cC}_{qu}^{(1,8)}]_{2332}$. We omit some coefficients which can be simply obtained from those listed by complex conjugation.}
    \label{tab:UVstates}
\end{table}

Inspecting this plot, we see most of the states generally induce sizeable $\kappa_c$ deviations, with the vectors $\mathcal{Q}_5$ and $\mathcal{Y}_5$ generating the largest $\Delta \kappa_c$ (with $\Delta \kappa_c \equiv \kappa_c -1$) for unit couplings. The octet scalar extension, $\Phi$, cannot couple to a colour-singlet current of quarks and therefore should not generate Yukawa interactions at one loop. In SMEFT language, the state populates the relevant coefficients at tree level such that one-loop $\kappa_c$ contributions cancel out, as is manifest in this plane. We therefore exclude this state from further study and discuss the remaining states below.

\subsubsection{Vectors $\mathcal{B}$ and $\mathcal{G}$}
Since the vector boson $\mathcal{G}$ (a colour octet) generates only $\mathcal{C}_{qu}^{(8)}$, which is less constrained than $\mathcal{C}_{qu}^{(1)}$ (see Sec.~\ref{eq:up_type}, particularly Fig.~\ref{fig:kappa_c_limits}), one would naively expect it to be a promising candidate for enhancing $\kappa_c$, with the converse holding for $\mathcal{B}$ (a gauge singlet). However, as seen in Tab.~\ref{tab:UVstates}, the couplings required to match to $[\hat{\cC}_{qu}^{(1,8)}]_{2332}$ also generate the $\Delta F=2$ coefficients $[\hat{\cC}_{qq}^{(1,3)}]_{2323}, [\hat{\cC}_{uu}]_{3232}$ and $[\hat{\cC}_{qu}^{(1,8)}]_{2323}$ at tree level, which face strong bounds from meson mixing observables \cite{Silvestrini:2018dos} and searches for same-sign top production. As an example, a $\mathcal{G}$ state with the couplings $(g^u_{\mathcal{G}})_{32}$ and $(g^q_{\mathcal{G}})_{23}$ set equal, faces particularly strong constraints from $\Delta m_{Bd}$. For a mass $M_{\mathcal{G}}=3$~TeV, current bounds already limit $|\Delta \kappa_c| \lesssim 2\times10^{-4}$, which is well below current and future direct sensitivity. Away from this equal coupling limit the constraint will weaken, however given that additional constraints exist from same-sign top production, we expect neither state is a good candidate for enhancing $\kappa_c$.

\subsubsection{Vectors $\mathcal{Q}_5$ and $\mathcal{Y}_5$}
The states $\mathcal Q_5$ and $\mathcal Y_5$ are coloured vector bosons.
$\mathcal{Q}_5 \sim (3,2,\frac56)$ arises in models of $SU(5)$ grand unification~\cite{PhysRevLett.32.438}, with both diquark and leptoquark couplings. It has also been studied in connection with $B$-anomalies~\cite{Cheung:2022zsb, Dorsner:2016wpm}. We remain agnostic about its UV origin but forbid its couplings to leptons to avoid baryon number violating interactions which would lead to fast proton decay. The vector diquark $\mathcal{Y}_5\sim (\bar{6},2,\frac56)$, realisable as a composite state, has been studied in relation to same-sign top production~\cite{Aguilar-Saavedra:2011iie, Zhang:2010kr, Degrande:2011rt, Kumar:2023zjj} and more broadly in~\cite{Assad:2017iib}. 
As seen in Fig.~\ref{fig:statesplot} and Tab.~\ref{tab:UVstates}, these vectors produce the largest deviations in $\kappa_c$ for unit couplings among the possible single-particle extensions, while only generating $\cC_{qu}^{(1,8)}$ at tree level.

We briefly discuss the broader flavour structures of the couplings involved, focussing on $\mathcal{Y}_5$ with similar conclusions holding for $\mathcal{Q}_5$, to determine under what assumptions $\kappa_c$ is the leading probe. The relevant Lagrangian terms are 
\begin{equation}
    \mathcal{L}_{\mathcal{Y}_5} \supset \frac{1}{2} (g_{\mathcal{Y}_5})_{ij} \, \mathcal{Y}_{5}^{AB\mu\dagger} \, \bar{u}_{Ri}^{(A|} \gamma_\mu i\sigma_2 q_{Lj}^{c|B)} + \text{h.c.}, 
\end{equation}
where the $(A|...|B)$ notation indicates a symmetric product of fundamental colour indices. Matching at tree level to the SMEFT gives \cite{deBlas:2017xtg}
\begin{align}
    [\hat{\cC}_{qu}^{(1)}]_{ijkl} = \frac{2(g_{\mathcal{Y}_5})^*_{lj}(g_{\mathcal{Y}_5})_{ki}}{3M_{\mathcal{Y}_5}^2}, ~~~ [\hat{\cC}_{qu}^{(8)}]_{ijkl} = \frac{(g_{\mathcal{Y}_5})^*_{lj}(g_{\mathcal{Y}_5})_{ki}}{M_{\mathcal{Y}_5 }^2}.
    \label{eq:y5_matching}
\end{align}
The flavour structure of $g_{\mathcal{Y}_5}$ is strongly constrained by the mass difference of neutral $D$-mesons ($\Delta m_D$); the product $(g_{\mathcal{Y}_5})_{11}(g_{\mathcal{Y}_5})_{22}$ contributes at tree level \cite{Zhang:2010kr}, and $(g_{\mathcal{Y}_5})_{31}(g_{\mathcal{Y}_5})_{32}$ at one loop (via $[\hat{\cC}_{qu}^{(1)}]_{1233}$ in the SMEFT). Keeping $M_{\mathcal{Y}_5}$ around the TeV scale therefore requires strongly suppressing couplings to the first quark generation. This suppression breaks the $U(2)$ flavour symmetry among the entries of $[\hat{\cC}_{qu}^{(1,8)}]_{ii33}$, thereby introducing strong constraints on $(g_{\mathcal{Y}_5})_{32}$ from $\varepsilon_{K}$ (as discussed in Sec.~\ref{sec:general_flavour}). Given the strong constraints on the parameter space, we limit our study to the minimal coupling texture which induces $\kappa_c$ modifications, allowing only $(g_{\mathcal{Y}_5})_{23}$ and $(g_{\mathcal{Y}_5})_{32}$ to be non-zero. 

\begin{figure}
    \centering
    \includegraphics[width=0.7\linewidth]{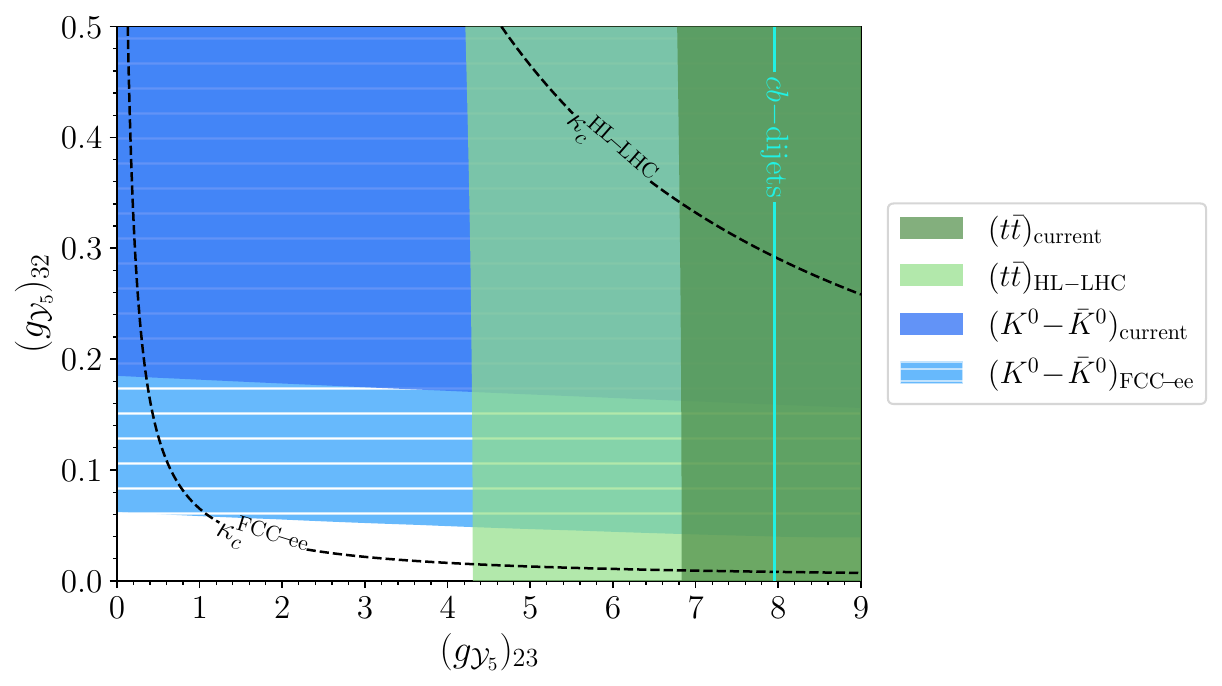}
    \caption{Parameter space for a $\mathcal{Y}_5$ with mass of 3 TeV, assuming the minimal couplings to enhance $\kappa_c$. Shaded regions are excluded at 95\% C.L. by current or projected measurements of the corresponding processes. Dashed lines indicate points in parameter space where $\kappa_c$ is enhanced to the experimental constraints in Tab.~\ref{tab:kappaprojections}.}
    \label{fig:y5_plot}
\end{figure}

Assuming $M_{\mathcal{Y}_5}=3$~TeV, we use Eq.~\eqref{eq:y5_matching} to map the SMEFT constraints from $t\bar{t}$ production and $K^0$--$\bar{K}^0$ mixing, calculated in Sec.~\ref{sec:general_flavour}, onto the couplings $(g_{\mathcal{Y}_5})_{23}$ and $(g_{\mathcal{Y}_5})_{32}$. A mass choice of 3 TeV puts it well beyond current sensitivity of searches for QCD-induced pair production \cite{Bordone:2021cca}.
ATLAS and CMS dijet resonance searches \cite{CMS:2018mgb, ATLAS:2019fgd, CMS:2019gwf, ATLAS:2018qto} constrain single production of these states; we use the recast provided in \cite{Bordone:2021cca} assuming decays to $cb$-dijets. 
The resulting parameter space is displayed in Fig.~\ref{fig:y5_plot}. We see that the coupling $(g_{\mathcal{Y}_5})_{32}$ is very strongly constrained by CP-violation in $K^0$--$\bar{K}^0$ mixing, mandating a large suppression with respect to $(g_{\mathcal{Y}_5})_{23}$. Modifications of $\kappa_c$ at HL-LHC sensitivity are excluded over the full parameter space. A small region of parameter space remains in which $\kappa_c$ can be enhanced to FCC-ee sensitivity, requiring a significant hierarchy between the couplings. 

We conclude that, due to the fact that these states populate operators involving vector currents of top quarks as well as the desired scalar current operators, upcoming $\kappa_c$ measurements will not generally be the leading probe, although there is a small region of parameter space to be tested at FCC-ee.

\subsubsection{Second Higgs doublet}
\label{sec:2hdm}

\begin{figure}[b]
    \centering
    \includegraphics[width=0.78\linewidth]{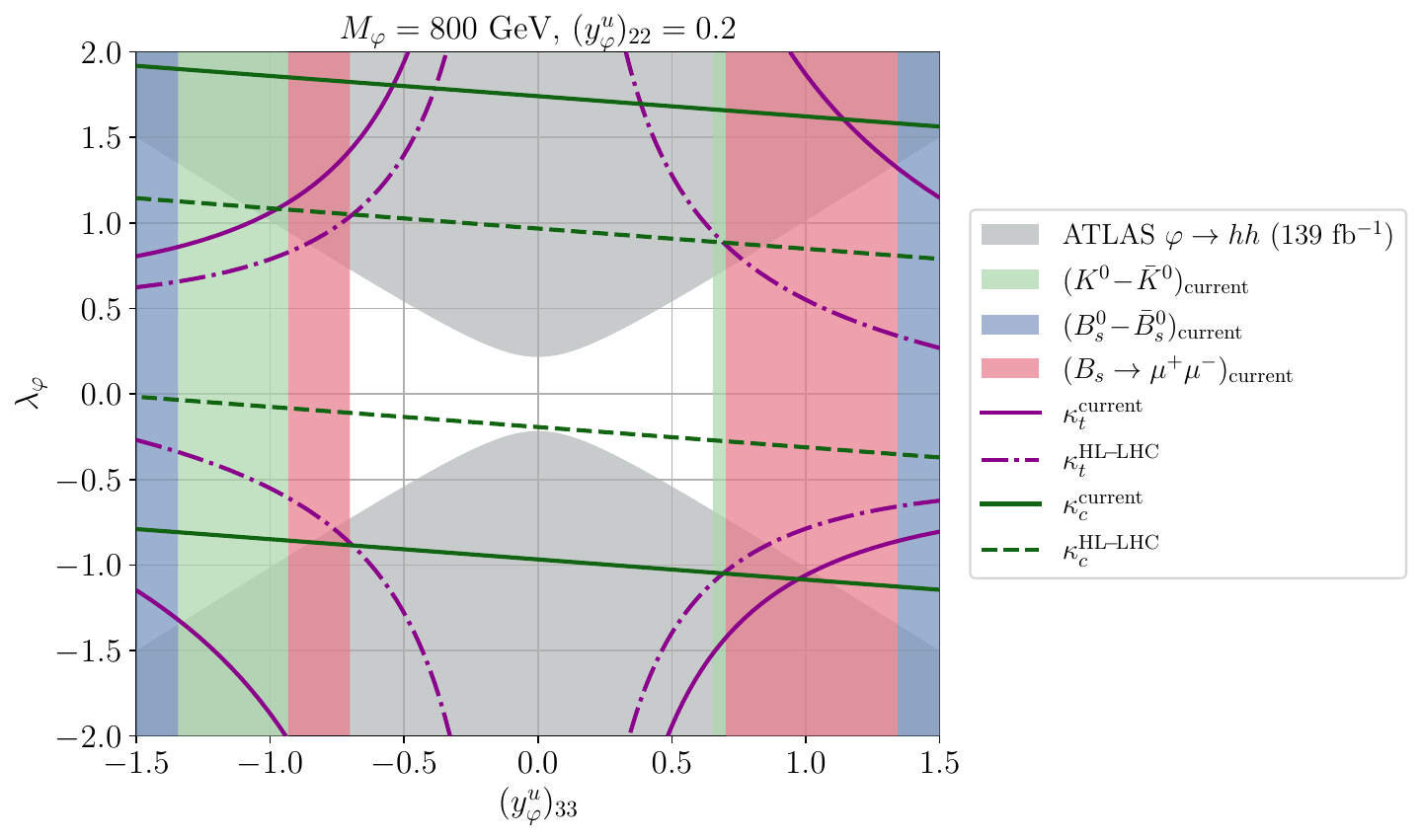}
    \caption{Parameter space of a second Higgs doublet with a mass of $800$ GeV, and coupling to charms of $(y_{\varphi}^u)_{22}=0.2$. We assume no coupling to the first generation, so $(y_{\varphi}^u)_{11}=0$. Current constraints and HL-LHC projections for $\kappa_c$ are shown with dark green lines. Purple lines show the same for $\kappa_t$. Flavour constraints, from meson mixing and $B_s\to \mu\mu$, bound the coupling to tops. Constraints from an ATLAS search for single production of a heavy neutral boson decaying to two Higgs bosons are shown in grey. All limits are at 95\% CL.}
    \label{fig:varphi}
\end{figure}

An additional heavy $SU(2)_L$ doublet scalar has been previously proposed in the literature as a mechanism for generating large light-quark Yukawa couplings, through mixing with the 125 GeV Higgs~\cite{Giannakopoulou:2024unn,Altmannshofer:2016zrn}. Here, we additionally consider its loop-level contributions through four-quark operators. The relevant interaction Lagrangian is given by
\begin{equation}
\label{eq:varphilag}
    -\mathcal{L}_\varphi \supset (y^{u}_{\varphi})_{ij} \, \varphi^\dagger i \sigma_2 \bar{q}_{Li}^T u_{Rj}
    +\lambda_\varphi\left(\varphi^\dagger H\right)\left(H^\dagger H\right) + \text{h.c}.
\end{equation}
We assume that $H$ has an otherwise SM-like potential, and $\lambda_\varphi$ is the only interaction between the two doublets. Note that there is no symmetry that could be imposed that would forbid the second term in the above Lagrangian while allowing the first (as well as the SM-like Yukawa terms). The second term leads to a mixing between the new doublet and the SM Higgs doublet, meaning that the 125 GeV Higgs inherits a small admixture of the neutral component of the new state. Assuming that this mixing is small (and that $\varphi$ has no other potential terms which would cause it to take a vev), matching at tree level to the SMEFT gives \cite{deBlas:2017xtg}
\begin{align}
\label{eq:varphimatching}
    [\hat{\cC}_{uH}]_{ij} = \frac{\lambda_{\varphi} ( y_{\varphi}^u)^*_{ji}}{M_{\varphi}^2}
    ,~
    [\hat{\cC}_{qu}^{(1)}]_{ijkl} = -\frac{(y_{\varphi}^u)^*_{li} (y_{\varphi}^u)_{kj} }{6 M_{\varphi}^2}
    ,~
    [\hat{\cC}_{qu}^{(8)}]_{ijkl} = -\frac{(y_{\varphi}^u)^*_{li} (y_{\varphi}^u)_{kj} }{ M_{\varphi}^2}
    ,~
    \cC_{\phi} = \frac{|\lambda_\varphi|^2}{M_{\varphi}^2},
\end{align}
from which (using \eqref{eq:rgecuh}) the full contribution to $\kappa_c$ is at first leading log:
\begin{equation}
    \kappa_c=1-\frac{v^3(y_{\varphi}^u)^*_{22}}{\sqrt{2}m_c M_\varphi^2}\left(\lambda_{\varphi}  -\frac{3 (y_{\varphi}^u)_{33}}{4\pi^2}(y_t^3-\lambda y_t)\log \left( \frac{m_h}{\Lambda}\right)\right),
\end{equation}
and similarly for $\kappa_u$ and $\kappa_t$, with appropriate substitutions. In Fig.~\ref{fig:varphi}, we show the corresponding contours of $\kappa_c$, for a fixed mass of $M_\varphi=800$ GeV and $(y_{\varphi}^u)_{22}=0.2$ (and assuming that $(y_{\varphi}^u)_{11}=0$). The grey region shows the $95\%$ CL exclusion from an ATLAS search for a heavy neutral Higgs decaying to a pair of SM Higgses~\cite{ATLAS:2022hwc}. The heavy $\varphi$ is singly-produced through its coupling to charm quarks, and then decays to pairs of tops, charms or Higgs bosons. The LHC production cross section $pp\to \varphi$ was calculated using the Mathematica implementation of the MMHT2014 PDFs~\cite{Harland-Lang:2014zoa}. The grey constraint weakens for increasing values of $(y_\varphi^u)_{33}$ because the branching ratio to $hh$ correspondingly decreases. This therefore allows the model to reach larger values of $\kappa_c$ than the situation where only tree level contributions are induced. However, large values of $|(y_\varphi^u)_{33}|\gtrsim 1$ are disfavoured by flavour, with the blue (light green) bounds from kaon ($B_s$) mixing calculated following Ref.~\cite{Crivellin:2013wna, Giannakopoulou:2024unn},\footnote{This calculation uses the full model rather than going via the SMEFT. Specifically, we include the pure charged-Higgs box and charged-Higgs-$W$/Goldstone box contributions to $\mathcal{C}_1$. Relative to the notation of \cite{Giannakopoulou:2024unn} we make the replacement $\mathcal{Y}^{H+}_{d_i \bar{u}_j} \to (V^T y^u_{\varphi})_{ij}$.} using the updated fit results in \cite{utfit}. The pink bounds from $\mathcal{B}(B_s\to \mu \mu)$ are calculated following Ref.~\cite{Crivellin:2019dun, Giannakopoulou:2024unn} with the SM input from \cite{Beneke:2017vpq, Beneke:2019slt} and the experimental average from \cite{Greljo:2022jac}. Given that it can decay to tops, $\varphi$ could also show up in LHC top pair resonance searches, however with no coupling to the first generation, the single production cross section is always too low to be within the sensitivity of current searches (such as~\cite{CMS:2025nqq}). 

\begin{figure}[t]
    \centering
    \includegraphics[width=0.7\linewidth]{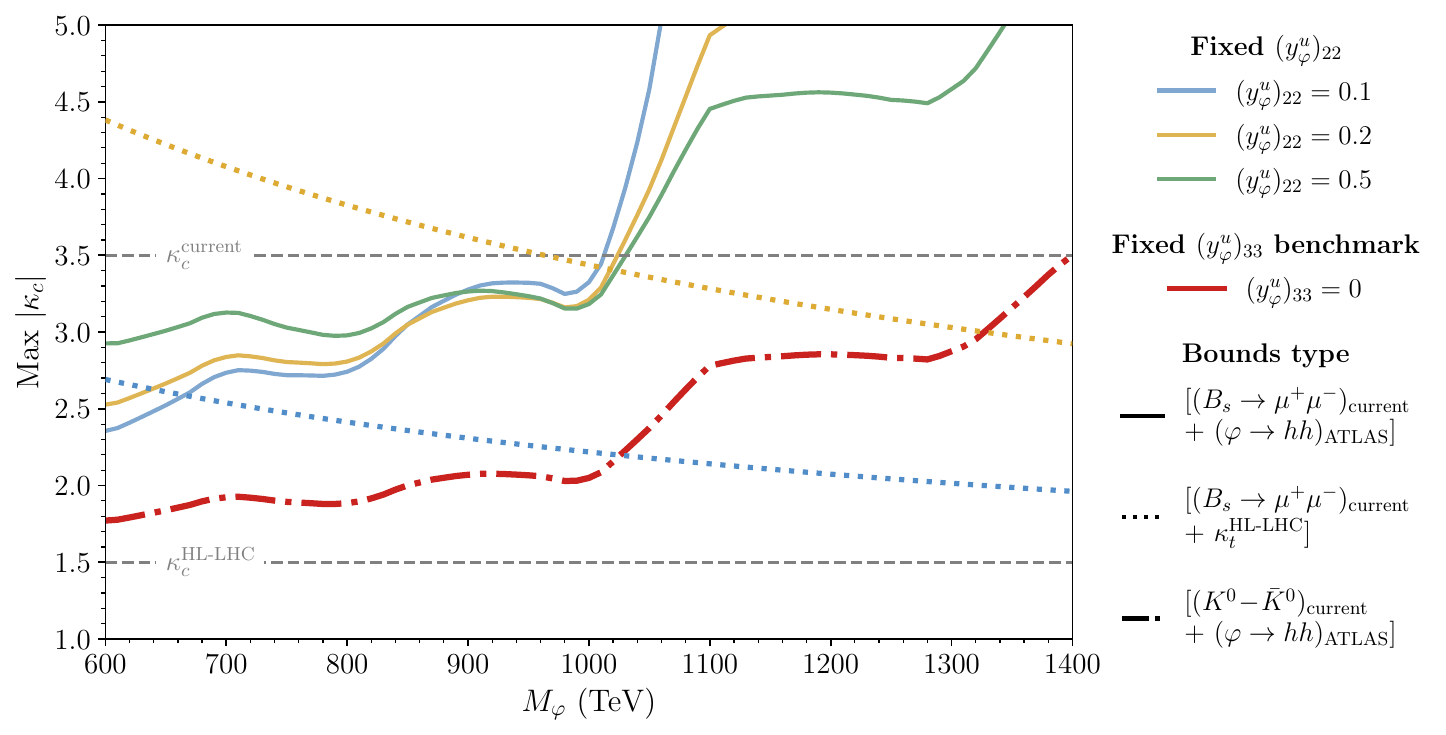}
    \caption{Maximum $|\kappa_c|$ enhancements allowed by correlated constraints for the $\varphi$ state under different mass and coupling assumptions. The solid blue, orange, and green lines show the maximum $|\kappa_c|$ consistent with current $B_s\to\mu^+\mu^-$ measurements and the ATLAS $\varphi\to hh$ search, while the dotted lines show the corresponding constraints when $\kappa_t$ is measured with the projected HL-LHC precision. In each of these cases, the top coupling $(y_\varphi^u)_{33}$ and $\lambda_{\varphi}$ are allowed to float to achieve the maximum $|\kappa_c|$ consistent with these constraints. The dark red line instead shows the maximum $|\kappa_c|$ allowed by current kaon mixing and ATLAS constraints when the top coupling is zero ($(y_{\varphi}^u)_{33}=0$), and the charm coupling $(y_\varphi^u)_{33}$ is allowed to float to maximise $|\kappa_c|$. The plot shows that larger values of $|\kappa_c|$ are achievable within models in which the top coupling is non-zero.}
    \label{fig:kappascan}
\end{figure}

It can be seen in Fig.~\ref{fig:varphi} that permitting a non-zero coupling to tops, $(y_\varphi^u)_{33}\neq 0$, allows for larger deviations in $\kappa_c$ than the case with only $\lambda_\varphi$ and $(y_\varphi^u)_{22}$ non-zero.
This feature is shown across a broader area of parameter space in Figure~\ref{fig:kappascan}, where the blue, orange and green solid and dotted lines show the maximum $|\kappa_c|$, for a given $(y_\varphi^u)_{22}$, allowed by the strongest current and HL-LHC constraints on $(y_\varphi^u)_{33}$ and $\lambda_\varphi$.\footnote{These constraints are significantly stronger than perturbativity bounds on the relevant couplings for the studied parameter space.} We show these alongside the benchmark with fixed $(y_\varphi^u)_{33}=0$, where the max $|\kappa_c|$ is instead set by the strongest current constraints on $(y_\varphi^u)_{22}$ and $\lambda_\varphi$.
Over the full mass range, the blue, orange and green solid lines have larger allowed $|\kappa_c|$ than the red dash-dotted benchmark. Assuming HL-LHC precision on $\kappa_t$, the dotted lines indicate that, for large masses and small $(y_\varphi^u)_{22}$, allowing non-zero $(y_\varphi^u)_{33}$ may reduce allowed deviations in $\kappa_c$. Nevertheless, a $\varphi$ state with modest couplings to both the first and second generations can induce large enhancements of $\kappa_c$ without violating correlated flavour and collider constraints. For masses above a TeV, current measurements of $\kappa_c$ already provide a strong probe of such a state.

We identify models involving this state as a key target and motivation for study of $\kappa_c$ at current and future colliders. Should a deviation be observed, searches for heavy Higgs decays, and/or correlated effects in flavour physics, can be used to discriminate between scenarios with and without couplings to the third generation.

\subsection{Single-particle completions for $\kappa_b$}

\begin{table}[]
    \centering
    \begin{tabular}{c|c|c|c|c|c}
    State & Spin & Quantum numbers & $[\check{\cC}_{quqd}^{(1)}]_{3333}$ & $[\check{\cC}_{quqd}^{(8)}]_{3333}$ & Other coeffs\\
    \hline
      $\varphi$ & $0$ & $(1,2,-\frac{1}{2})$ & $-\frac{(y_\varphi^u)_{33}(y_\varphi^d)^*_{33}}{M_\varphi^2}$ & $0$ & $[\check{\cC}_{qu}^{(1,8)}]_{3333}$, $[\check{\cC}_{qd}^{(1,8)}]_{3333}$ \\
      $\Phi$ & $0$ & $(8,2,-\frac{1}{2})$ & $0$ & $-\frac{(y_\Phi^{dq})^*_{33}(y_\Phi^{qu})_{33}}{M_\Phi^2}$ & $[\check{\cC}_{qu}^{(1,8)}]_{3333}$, $[\check{\cC}_{qd}^{(1,8)}]_{3333}$ \\
      $\omega_1$ & $0$ & $(3,1,-\frac{1}{3})$ & $\frac{4(y_{\omega_1}^{qq})_{33}(y_{\omega_1}^{ud})^*_{33}}{3M_{\omega_1}^2}$ & $-\frac{4(y_{\omega_1}^{qq})_{33}(y_{\omega_1}^{ud})^*_{33}}{M_{\omega_1}^2}$ & $[\check{\cC}_{ud}^{(1,8)}]_{3333}$, $[\check{\cC}_{qq}^{(1,3)}]_{3333}$ \\
    \end{tabular}
    \caption{States generating $[\check{\cC}_{quqd}^{(1,8)}]_{3333}$ at tree level, and their tree level matching, from Ref.~\cite{deBlas:2017xtg}. The final column gives other coefficients and flavour structures that are inevitably generated by the same couplings which generate $[\check{\cC}_{quqd}^{(1,8)}]_{3333}$.}
    \label{tab:UVstateskappab}
\end{table}

\begin{figure}
    \centering
    \includegraphics[width=0.6\linewidth]{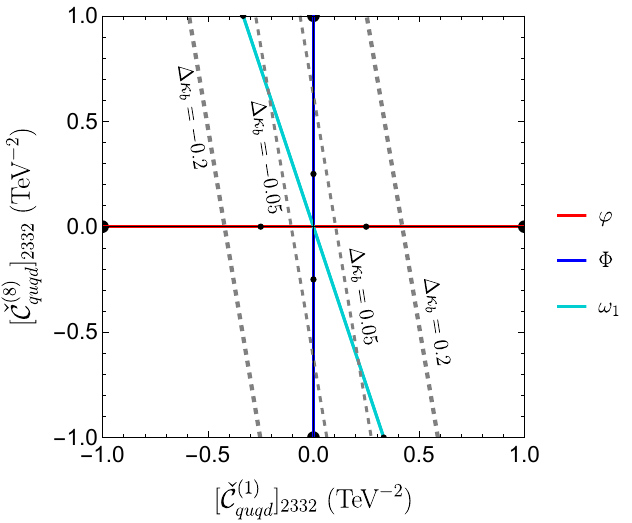}
    \caption{Plot depicting the matching of the three states listed in Tab.~\ref{tab:UVstateskappab} onto the coefficients $[\check{\cC}_{quqd}^{(1,8)}]_{3333}$. The larger (smaller) black blobs on the lines show the points for which the mass of the state is at $1$ TeV ($2$ TeV), assuming couplings of unit magnitude. Contours of $\Delta \kappa_b$ are also shown (with $\Delta \kappa_b \equiv \kappa_b -1$), assuming running from a scale of $\Lambda=2$ TeV.}
    \label{fig:statesplotkappab}
\end{figure}

We gather all the states which match to $[\cC_{quqd}^{(1,8)}]_{3333}$ at tree level in Tab.~\ref{tab:UVstateskappab}, noting also the other coefficients which are inevitably generated by the relevant couplings.\footnote{The colour sextet $\Omega_1$ also matches to $\cC^{(1)/(8)}_{quqd}$ in a similar way to the colour triplet $\omega_1$, however the gauge structure of $\Omega_1$ enforces antisymmetry in the flavour indices of its coupling to quark doublets. This means that it cannot generate $[\cC_{quqd}^{(1,8)}]_{3333}$.} These are all scalar particles. In Fig.~\ref{fig:statesplotkappab}, we show the matching relations for each state in the plane of the relevant coefficients, alongside contours of $\kappa_b$ and benchmark enhancements for unit couplings. 

As seen in Fig.~\ref{fig:kappa_b}, new physics generating $[\check{\cC}_{quqd}^{(8)}]_{3333}$ alone cannot induce deviations in $\kappa_b$ observable at HL-LHC, due to constraints from $B\to X_s \gamma$. We can therefore discard the colour octet scalar $\Phi$ as an option for HL-LHC-level $\kappa_b$ deviations. The remaining states are the doublet $\varphi$, and the scalar colour triplet diquark $\omega_1$. The second Higgs doublet $\varphi$ generates $\kappa_b$ at both tree and loop level, in an analogous way to the mechanism for $\kappa_c$ discussed in Sec.~\ref{sec:2hdm}. We do not repeat the analysis for the case of $\kappa_b$, however we comment that direct production constraints will be weaker in the case that the $\varphi$ only couples to $b$ quarks compared to charm quarks, due to the additional PDF suppression.

\begin{figure}
    \centering
    \includegraphics[width=0.85\linewidth]{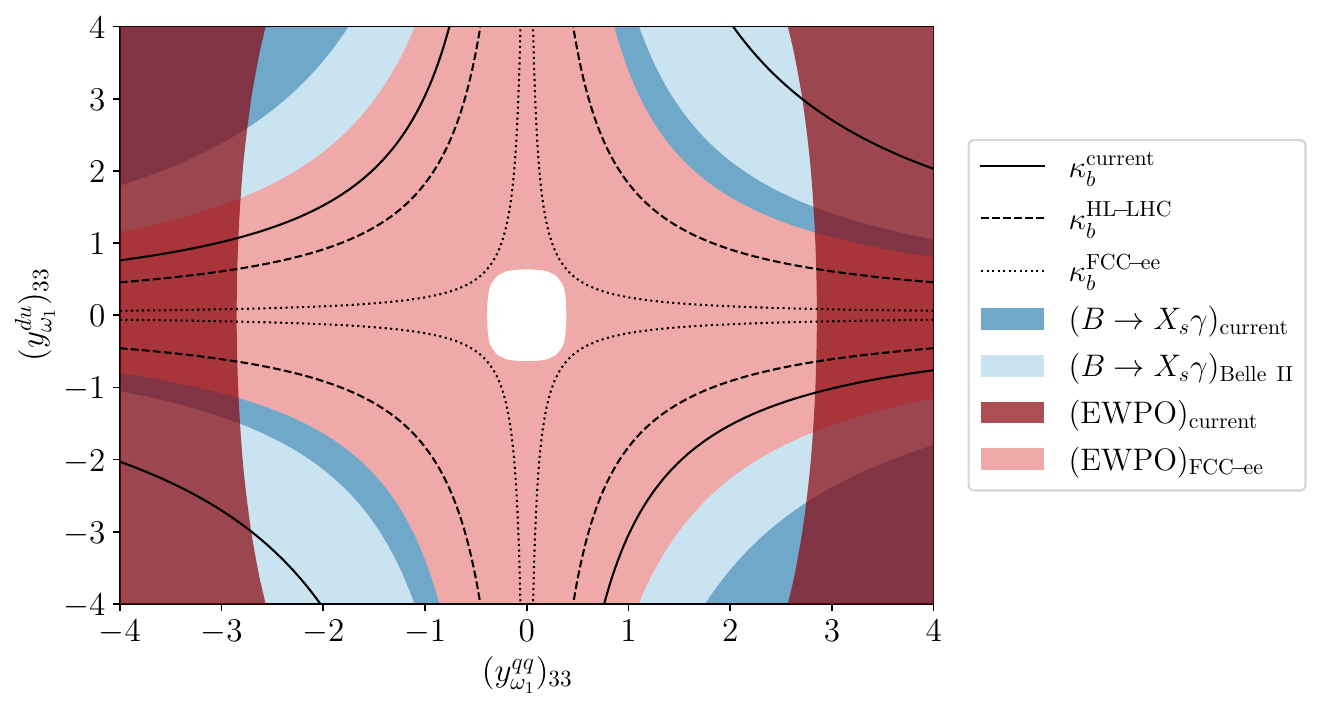}
    \caption{Experimental constraints and future sensitivities at 95\% CL on the parameter space of the couplings (assumed real) of the triplet scalar diquark $\omega_1$, with a mass of 3 TeV. Current constraints from EWPO fits and $\mathcal{B}(B\to X_s \gamma)$ are shown in dark red and blue respectively, while future sensitivities are shown in paler colours. Contours of $\kappa_b$ sensitivity are shown with black lines, for current (solid), HL-LHC (dashed) and FCC-ee (dotted). }
    \label{fig:omega1}
\end{figure}

For the $\omega_1$ diquark, current constraints and future sensitivity in the plane of the couplings $(y_{\omega_1}^{ud})_{33}$ and $(y_{\omega_1}^{qq})_{33}$ are shown in Fig.~\ref{fig:omega1}. This plot assumes a mass of 3 TeV for the diquark, putting it well above the range of direct pair production bounds. Current bounds on a pair produced scalar triplet diquark decaying to dijets are around 1 TeV~\cite{Englert:2024nlj, CMS:2022wyd}. Single production is also possible through its quark couplings, but if it only couples to third generation quarks, PDF suppression renders the cross section too small to be subject to constraints. 

Constraints from electroweak precision observables arise because at one-loop, $[\cC_{ud}^{(1)}]_{3333}$ and $[\cC_{qq}^{(1)}]_{3333}$ contribute to $Z\to \bar b b$, while at two loop $[\cC_{qq}^{(1)}]_{3333}$ gives contributions to the $T$ parameter~\cite{Allwicher:2023aql}. We evaluate these contributions using recent two-loop calculations of the observables in terms of SMEFT Wilson coefficients~\cite{Haisch:2024wnw},
and find constraints by fitting to the full set of $Z$ pole observables from LEP and SLD~\cite{ALEPH:2005ab,Janot:2019oyi}, as well as the latest average of the $W$ mass~\cite{LHC-TeVMWWorkingGroup:2023zkn}. Constraints from current and future Belle II sensitivity to $\mathcal{B}(B\to X_s\gamma)$ are shown in darker and lighter blue, respectively. In some parts of the parameter space, current constraints on $\kappa_b$ provide the strongest constraint on this model, and HL-LHC measurements of $\kappa_b$ will explore new parameter space. 

At a Tera-Z run at FCC-ee, the huge number of $Z$ and $W$ bosons produced will enormously improve on the sensitivity of electroweak precision observables. The salmon region shows projected sensitivity taking experimental projections from Ref.~\cite{Selvaggi:2025kmd} and assuming the `aggressive' scenario for theoretical errors described in Ref.~\cite{deBlas:2025gyz}. It can be seen that if a deviation in $\kappa_b$ observed at FCC-ee is due to this model, then signals should also be seen in electroweak precision tests.

The parameter space shown in Fig.~\ref{fig:omega1} represents the minimal couplings needed to generate $\kappa_b$ in this model. Introducing couplings to any other down-type quark will lead to strong constraints from $B_{s,d}$-meson mixing which will be generated at one loop (see e.g.~\cite{Giudice:2011ak}). We do not quantify these here, but simply note that this model will need to be aligned along the third generation of down-type quarks.

\section{Conclusions}
\label{sec:concs}
If a deviation in the magnitude of Higgs couplings to charm or bottom quarks is measured at HL-LHC, it could be the first sign of new physics generating four-quark operators involving scalar currents (of top quarks and of charm or bottom quarks, respectively). In the case of Higgs couplings to up, down and strange quarks, constraints from $\bar t t$ pair production and/or radiative $B$ decays already preclude the possibility of four-quark operators generating Higgs Yukawa deviations that could be detected at HL-LHC, although there may still be room for detection at FCC-ee (in some cases depending on the outcome of HL-LHC top pair measurements). 

The scalar current operators which can radiatively generate Higgs-quark coupling shifts are necessarily flavour-breaking, and are not included within common flavour-symmetric subsets of operators. Hence their phenomenology is relatively unexplored, and our analysis provides some of the first bounds on these operators. Their flavour-breaking nature leads to effects in flavour physics (particularly meson mixing and $B\to X_s \gamma$), but conversely suppresses their effects in some flavour conserving processes, such as $t\bar t$ production (where they do not interfere with the SM), and electroweak precision tests (where their amplitudes are suppressed by powers of $m_q$/$v$). Allowing a broader flavour structure on the operators can therefore lead to stronger constraints, especially if it relates scalar and vector currents. Nevertheless, we find that a third-generation-philic structure involving only scalar currents (for example as induced by a 2HDM with quark couplings) could first show up in $\kappa_c$ or $\kappa_b$ deviations.

Going beyond the effective field theory, we also investigated simple tree-level UV completions of the relevant operators, including a second Higgs doublet and scalar and vector diquarks. We find that a second Higgs doublet with couplings to top quarks as well as to charm quarks can induce a larger deviation in $\kappa_c$ than the case where only the minimal charm coupling is present, given current direct and indirect constraints. For $\kappa_b$, a scalar diquark coupled only to third generation quarks can produce deviations observable at HL-LHC, while at FCC-ee it will be tested both by $\kappa_b$ and by electroweak precision on the $Z$ pole.

This work demonstrates that future improvements in constraining quark Yukawas can tell us about new physics without direct couplings to the Higgs. Measurements of quark Yukawas at the LHC and future colliders can offer some of the strongest tests of four-quark operators involving a scalar current of tops, which can be generated by a variety of new states at the TeV-scale. 

\appendix
\section{Formulae for flavour observables}
\subsection{$B_s\to \mu \mu$}
\label{app:bsmumu_formulae}
Predictions for $b\to s \mu\mu$ decay observables are traditionally written in terms of the coefficients of the effective Lagrangian
\begin{equation}
	\mathcal{L_{\text{eff}}} \supset -\frac{4 G_F}{\sqrt{2}} \frac{\alpha}{4 \pi} V_{ts}^* V_{tb} \sum_i \mathcal{C}_i \Op_{i}, \label{eq:bsll}
\end{equation}
with (including here only the operators relevant for our analysis)
\begin{align}
	\Op_9^{\ell} = (\bar s \gamma_\alpha P_{L (R)} b) (\bar \ell \gamma^\alpha \ell), ~~~
	\Op_{10}^{\ell} = (\bar s \gamma_\alpha P_{L (R)} b) (\bar \ell \gamma^\alpha \gamma_5 \ell).
\end{align}
The theory prediction for $\mathcal{B}(B_s\to \mu^+ \mu ^-)$ is then written as 
\begin{equation}
    \mathcal{B}(B_s\to \mu^+ \mu ^-) = \mathcal{B}(B_s\to \mu^+ \mu ^-)_{\text{SM}}\left|1+\frac{\cC_{10}}{\cC_{10}^{\text{SM}}}\right|^2,
\end{equation}
where we use $\cC_{10}^\text{SM}=-4.188$ \cite{Allwicher:2023shc}. For this analysis, we take the SM prediction from \cite{Czaja:2024the} together with the experimental average from \cite{ParticleDataGroup:2026aaa}: 
\begin{equation}
    \mathcal{B}(B_s \to \mu^+ \mu^-)_{\text{SM}} = (3.64 \pm 0.12) \times10^{-9},~~~
    \mathcal{B}(B_s \to \mu^+ \mu^-)_{\text{exp}} = (3.34\pm0.27) \times10^{-9}.
\end{equation}
For HL-LHC projections we assume an experimental uncertainty $\pm 0.16 \times10^{-9}$ following \cite{ATLAS:2025lrr}.

\subsection{$B\to X_{s/d}\gamma$}
\label{app:bXdgamma_formulae}
Predictions for the branching ratios $\mathcal{B}(B\to X_{s/d} \gamma)$ are conventionally written in terms of the effective Lagrangian
\begin{align}
    \mathcal{L_{\text{eff}}} \supset \frac{4G_F }{\sqrt{2}} V_{tb}V_{ts}^*\frac{m_b}{16\pi^2}\left(e\cC_{7(')}^{bs}(\overline{s}\, \sigma^{\mu\nu} P_{R(L)} b)F_{\mu\nu} +g_s \cC_{8(')}^{bs}(\overline{s}\, \sigma^{\mu\nu}T^A P_{R(L)} b)G_{\mu\nu}^A\right. \nonumber\\
    \left. +e\cC_{7(')}^{bd}(\overline{d}\, \sigma^{\mu\nu} P_{R(L)} b)F_{\mu\nu} +g_s \cC_{8(')}^{bd}(\overline{d}\, \sigma^{\mu\nu}T^A P_{R(L)} b)G_{\mu\nu}^A\right),\label{eq:bxdgamma}
\end{align}
where $F_{\mu\nu}$ and $G^A_{\mu\nu}$ are the field strength tensors of QED and QCD, $T^A$ are the $SU(3)_c$ generators and coefficients are defined at $\mu=m_b$. Including all SM contributions in the constant terms (so the Wilson coefficients here mean only the BSM contributions), the branching ratio for $B \to X_s \gamma$ is written \cite{Bruggisser:2021duo}\footnote{During the completion of this work, a new calculation of $\mathcal{B}(B \to X_s \gamma)$ became available~\cite{Misiak:2026sqy}. Since their formula for the branching ratio does not include quadratic BSM terms, we stick with the formula of \cite{Bruggisser:2021duo} here, for consistency with other parts of our analysis.}
\begin{align}
\mathcal{B}(B \to X_s \gamma)\times 10^{4}
=
&\left[3.26- 15.17\,\mathcal{C}_7^{bs}- 0.77\,\mathcal{C}_8^{bs}+ 1.66\,\mathcal{C}_7^{bs} \mathcal{C}_8^{bs}+ 1.36\,\mathcal{C}_7^{bs} \mathcal{C}_{8'}^{bs}\right. \nonumber\\
&\left. + 18.03\,\left[(\mathcal{C}_7^{bs})^2 + (\mathcal{C}_{7'}^{bs})^2\right] + 0.20\,(\mathcal{C}_8^{bs})^2 + 0.09\,(\mathcal{C}_{8'}^{bs})^2 \right],
\end{align}
where we assume 5\% theory uncertainty \cite{Misiak:2020vlo}. The branching ratio $B \to X_d \gamma$ is given by \cite{Bause:2022rrs}
\begin{align}
\mathcal{B}(B \to X_d \gamma)\times 10^{5}
=&\left[
1.77 - 6.17\,\mathcal{C}_7^{bd}- 0.28\,\mathcal{C}_8^{bd} + 7.66\,\left[(\mathcal{C}_7^{bd})^2 + (\mathcal{C}_{7'}^{bd})^2\right]\right. \nonumber\\
& \left. + 0.28\,\left[(\mathcal{C}_8^{bd})^2 + (\mathcal{C}_{8'}^{bd})^2\right] + 0.53\,\left(\mathcal{C}_7^{bd} \mathcal{C}_8^{bd} + \mathcal{C}_{7'}^{bd} \mathcal{C}_{8'}^{bd}\right)\right],
\end{align}
with a theory uncertainty of 15\%. 
For experimental inputs, we take the current world average \cite{HFLAV:2024ctg}
\begin{equation}
    \mathcal{B}(B \to X_s \gamma)^{\text{exp}}= (3.49\pm 0.19)\times10^{-4}
\end{equation}
and the extrapolated BaBar measurement \cite{BaBar:2010vgu, Misiak:2015xwa} 
\begin{equation}
    \mathcal{B}(B \to X_d \gamma)^{\text{exp}}= (1.41\pm 0.57)\times10^{-5},
\end{equation}
alongside the 50~ab$^{-1}$ Belle~II projections from \cite{ATLAS:2025lrr} and \cite{Belle-II:2018jsg} respectively.

\subsection{$\Delta F=2$}
\label{app:meson_mixing}
The relevant effective Lagrangian for meson mixing is
\begin{equation}
    \mathcal{L_\text{eff}}\supset -\sum_{i=1}^{5} \mathcal{C}_i \Op_i -\sum_{i=1}^{3}\tilde{\mathcal{C}}_{i} \tilde{\Op}_i,
\end{equation}
where for $K^0 - \bar{K}^0$ mixing the relevant structures are
\begin{align}
\mathcal{O}^{sd}_1 &= \left(\bar{s}_a \gamma_\mu P_L d_a\right)\left(\bar{s}_b \gamma^\mu P_L d_b\right), \\
\mathcal{O}^{sd}_2 &= \left(\bar{s}_a P_L d_a\right)\left(\bar{s}_b P_L d_b\right), \\
\mathcal{O}^{sd}_3 &= \left(\bar{s}_a P_L d_b\right)\left(\bar{s}_b P_L d_a\right), \\
\mathcal{O}^{sd}_4 &= \left(\bar{s}_a P_L d_a\right)\left(\bar{s}_b P_R d_b\right), \\
\mathcal{O}^{sd}_5 &= \left(\bar{s}_a P_L d_b\right)\left(\bar{s}_b P_R d_a\right),
\end{align}
where $a,b$ are colour indices, alongside $\tilde{\Op}_{1,2,3}$, which are obtained from $\Op_{1,2,3}$ by the exchange $P_L\leftrightarrow P_R$. The corresponding operators for $D^0 - \bar{D}^0$ and $B_d^0 - \bar{B}_d^0$ and $B_s^0 - \bar{B}_s^0$ are obtained by appropriate flavour replacements.

For mixing observables in the down-type sector, the relevant tree level matching relations to SMEFT operators are \cite{Harnik:2012pb, Silvestrini:2018dos, Aebischer:2015fzz}
\begin{align}
\mathcal{C}^{ij}_1 &= -[\check{\mathcal{C}}_{qq}^{(1)}]_{ijij} - [\check{\mathcal{C}}_{qq}^{(3)}]_{ijij},\\
\tilde{\mathcal{C}}^{ij}_1 &= -[\check{\mathcal{C}}_{dd}]_{ijij},\\
\mathcal{C}^{ij}_2 &= -\frac{v^4}{4m_h^2}([\check{\mathcal{C}}_{dH}]_{ij})^2\\
    \tilde{\mathcal{C}}^{ij}_2 &= -\frac{v^4}{4m_h^2}([\check{\mathcal{C}}^*_{dH}]_{ji})^2\\
\mathcal{C}^{ij}_4 &= [\check{\mathcal{C}}_{qd}^{(8)}]_{ijij} -\frac{v^4}{2m_h^2}([\check{\mathcal{C}}^*_{dH}]_{ji}[\check{\mathcal{C}}_{dH}]_{ij}),\\
\mathcal{C}^{ij}_5 &= 2[\check{\mathcal{C}}_{qd}^{(1)}]_{ijij} - \frac{1}{3}[\check{\mathcal{C}}_{qd}^{(8)}]_{ijij},
\end{align}
where all coefficients are evaluated at the electroweak scale and contributions from Yukawa operators are formally a $\Op(\frac{1}{\Lambda^4})$ effect. Relations for coefficients related to $D^0 - \bar{D}^0$ mixing are obtained by replacing the $SU(2)$ singlet $d$ with $u$ and assuming an up-aligned basis for the SMEFT coefficients.

For constraints on the real and imaginary parts of these low-energy coefficients, we use the current lower bounds on the effective new physics scales $\Lambda = \frac{1}{\sqrt{\mathcal{C}_i}}$ given in \cite{utfit}, where their analysis follows \cite{UTfit:2007eik}. For FCC-ee projected bounds on these coefficients, we account for improvements in determinations of $V_{cb}$ \cite{Marzocca:2024mkc, Liang:2024hox} with a roughly factor of three increase in the effective new physics scale \cite{deBlas:2025gyz}. To propagate these bounds onto the UV-scale coefficients of interest, we RGE evolve SMEFT coefficients to the electroweak scale and match to the relevant low-energy coefficients, analogously to \cite{Silvestrini:2018dos}. 

\section{Constraints on all-third-generation coefficients}
\label{app:thirdgen}
We report here the individual 95\% CL constraints on studied Wilson coefficients restricted to third generation flavour indices. Tabs.~\ref{tab:qu_third_gen_flavour_constraints} and \ref{tab:quqd_third_gen_flavour_constraints} summarise the constraints from flavour observables calculated in Sec.~\ref{sec:general_flavour} (using the formulae in Apps.~\ref{app:bsmumu_formulae} and \ref{app:meson_mixing}) and Sec.~\ref{sec:down_type_eft} (using App.~\ref{app:bXdgamma_formulae}), respectively, presented in both the up- and down-aligned Yukawa bases. In Tab.~\ref{tab:qu_third_gen_collider_constraints} we summarise further constraints from EWPOs and four-top production, calculated as in Sec.~\ref{sec:general_flavour}. The flavour and EWPO constraints are given assuming that the coefficients are evaluated at a matching scale $\Lambda=3$~TeV, with renormalisation group running down to the scale of the observables included in the calculation, while running effects are neglected for the four-top bounds. 

\begin{table}[t]
\centering
\scalebox{0.95}{
\begin{tabular}{|c|c|c|c|c|c|c|}
\hline
& \multicolumn{2}{c|}{$K^0$--$\bar K^0$ mixing}
& \multicolumn{2}{c|}{$B_s$--$\bar B_s$ mixing}
& \multicolumn{2}{c|}{$B_s\to\mu^+\mu^-$} \\
\cline{2-7}
& Current & FCC-ee
& Current & FCC-ee
& Current & HL-LHC \\
\hline
$[\hat{\mathcal{C}}_{qu}^{(1)}]_{3333}$ & [-1.2, 1.2] & [-0.11, 0.11] & [-0.68, 0.68] & [-0.062, 0.062] & [-0.078, 0.27] & [-0.11, 0.12] \\
\hline
$[\check{\mathcal{C}}_{qu}^{(1)}]_{3333}$ & [-180, 180] & [-17, 17] & [-11, 11] & [-1.0, 1.0] & [-4.7, 1.4] & [-2.0, 1.9] \\
\hline
$[\hat{\mathcal{C}}_{qu}^{(8)}]_{3333}$ & [-3.0, 3.0] & [-0.29, 0.29] & [-1.7, 1.7] & [-0.16, 0.16] & [-6.5, 22] & [-9.1, 10] \\
\hline
$[\check{\mathcal{C}}_{qu}^{(8)}]_{3333}$ & [-210, 210] & [-21, 21] & [-16, 16] & [-1.5, 1.5] &[-21, 6.0] & [-8.8, 8.4]  \\
\hline
\end{tabular}}
\caption{Individual 95\% CL limits on the Wilson coefficients from flavour observables.}
\label{tab:qu_third_gen_flavour_constraints}
\end{table}

\begin{table}[t]
\centering
\begin{tabular}{|c|c|c|}
\hline
& \multicolumn{2}{c|}{$B\to X_s \gamma$} \\
\cline{2-3}
& Current & Belle II \\
\hline
$[\hat{\mathcal{C}}_{quqd}^{(1)}]_{3333}$ & [-0.016, 0.0055] & [-0.010, 0.0094] \\
\hline
$[\check{\mathcal{C}}_{quqd}^{(1)}]_{3333}$ & [-0.061, 0.18] & [-0.10, 0.11] \\
\hline
$[\hat{\mathcal{C}}_{quqd}^{(8)}]_{3333}$ & [-0.019, 0.0064] & [-0.012, 0.011] \\
\hline
$[\check{\mathcal{C}}_{quqd}^{(8)}]_{3333}$ & [-0.15, 0.44] & [-0.26, 0.28] \\
\hline
\end{tabular}
\caption{Individual 95\% CL limits on the Wilson coefficients from flavour observables.}
\label{tab:quqd_third_gen_flavour_constraints}
\end{table}

\begin{table}[t]
\centering
\begin{tabular}{|c|c|c|c|c|}
\hline
& \multicolumn{2}{c|}{EWPO}
& \multicolumn{2}{c|}{$pp \to t\bar t t \bar t$} \\
\cline{2-5}
& Current & FCC-ee
& Current & HL-LHC \\
\hline
$[\mathcal{C}_{qu}^{(1)}]_{3333}$ & [-0.15, 0.53] & [-0.0075, 0.0075] & [-3.2, 2.7] & [-2.2, 1.8] \\
\hline
$[\mathcal{C}_{qu}^{(8)}]_{3333}$ & [-5.3, 23] & [-0.32, 0.32] & [-5.6, 6.5] & [-3.6, 4.5] \\
\hline
\end{tabular}
\caption{Individual 95\% CL limits on Wilson coefficients from electroweak precision observables and four-top production.}
\label{tab:qu_third_gen_collider_constraints}
\end{table}

\section{Top pair production}
\label{app:toppair}
In this Appendix we present partonic tree-level top pair production cross sections in the presence of the operator coefficients $[\cC_{qu}^{(1,8)}]_{i33i}$,  $[\cC_{qu}^{(1,8)}]_{3ii3}$, $[\cC_{quqd}^{(1,8)}]_{j33j}$ and $[\cC_{quqd}^{(1,8)}]_{33jj}$ (for $i=1,2$, $j=1,2,3$), which to our knowledge have not appeared in previous literature. The contributions of these operators do not interfere with the SM, since they involve a scalar (or tensor) current of initial state quarks. Their first effects are thus at $O(\Lambda^{-4})$ in the SMEFT expansion. These pieces can nevertheless lead to important constraints. 

\subsection{Amplitudes and partonic cross section for $u_i\bar u_i \to t\bar t$ via $\cC_{qu}^{(1,8)}$}
The relevant SMEFT tree-level amplitudes for the process $u_i^b (p_1) \bar u_i^a(p_2) \to t^c(p_3)\bar t^d(p_4)$ are:
\begin{align}
    i \mathcal{A}_{VLR}&=iC_{VLR}\left[\bar{u}(p_3)^c \gamma^{\mu}P_Lu(p_1)^b\right]\left[\bar{v}(p_2)^a \gamma_{\mu}P_Rv(p_4)^d\right],\\
    i \mathcal{A}_{VRL}&=iC_{VRL}\left[\bar{u}(p_3)^c\gamma^{\mu}P_Ru(p_1)^b\right]\left[\bar{v}(p_2)^a \gamma_{\mu}P_Lv(p_4)^d\right],
\end{align}
where $a,b,c,d$ are colour indices and
\begin{align}
    C_{VLR}&=-[\cC_{qu}^{(1)}]_{i33i}\,\delta_{cb}\delta_{ad}+\frac{1}{6}[\cC_{qu}^{(8)}]_{i33i}(3\delta_{cd}\delta_{ab}-\delta_{cb}\delta_{ad}),\\
    C_{VRL}&=-[\cC_{qu}^{(1)}]_{3ii3}\,\delta_{cb}\delta_{ad}+\frac{1}{6}[\cC_{qu}^{(8)}]_{i33i}(3\delta_{cd}\delta_{ab}-\delta_{cb}\delta_{ad}).
\end{align}
The (spin and colour summed and averaged) squared matrix elements are
\begin{align}
|\overline{\mathcal{A}_{VLR}}|^2&=\hat{s}^2\left(1-\frac{2m_t^2}{\hat{s}} \right)\left[\left([\cC_{qu}^{(1)}]_{i33i}\right)^2+\frac{2}{9}\left([\cC_{qu}^{(8)}]_{i33i}\right)^2\right],\\
|\overline{\mathcal{A}_{VRL}}|^2&=\hat{s}^2\left(1-\frac{2m_t^2}{\hat{s}} \right)\left[\left([\cC_{qu}^{(1)}]_{3ii3}\right)^2+\frac{2}{9}\left([\cC_{qu}^{(8)}]_{3ii3}\right)^2\right].
\end{align}
The interference terms between the two amplitudes, as well as between each amplitude and the SM, are zero.
Then the total partonic cross section is given by
\begin{align}
    \hat{\sigma}&=\frac{\hat s}{16\pi}\left(1-\frac{2m_t^2}{\hat{s}} \right)\sqrt{1-\frac{4m_t^2}{\hat s}}\nonumber \\
&\times\left[\left([\cC_{qu}^{(1)}]_{i33i}\right)^2+\left([\cC_{qu}^{(1)}]_{3ii3}\right)^2+\frac{2}{9}\left([\cC_{qu}^{(8)}]_{i33i}\right)^2+\frac{2}{9}\left([\cC_{qu}^{(8)}]_{3ii3}\right)^2 \right].
\end{align}
This can be compared to the situation in which the flavour indices are arranged such that the operator generates a \emph{vector current} of top quarks, i.e.~$[\cC_{qu}^{(1,8)}]_{ii33}$ or $[\cC_{qu}^{(1,8)}]_{33ii}$. In this case, the operator(s) with colour octet currents has a tree-level interference term with QCD, while the operator(s) with colour singlet currents does not interfere at leading order with the QCD process (though it does with electroweak production, which we neglect here). The relevant SMEFT tree-level amplitudes are now
\begin{align}
    i \mathcal{A}_{VRR}&=iC_{VRR}\left[\bar{v}(p_2)^a \gamma^{\mu}P_Lu(p_1)^b\right]\left[\bar{u}(p_3)^c \gamma_{\mu}P_Rv(p_4)^d\right],\\
    i \mathcal{A}_{VLL}&=iC_{VLL}\left[\bar{v}(p_2)^a \gamma^{\mu}P_Ru(p_1)^b\right]\left[\bar{u}(p_3)^c \gamma_{\mu}P_Lv(p_3)^d\right],
\end{align}
where $a,b,c,d$ are colour indices and
\begin{align}
    C_{VRR}&=[\cC_{qu}^{(1)}]_{ii33}\delta_{ab}\delta_{cd}+[\cC_{qu}^{(8)}]_{ii33}T^A_{ab}T^A_{cd},\\
    C_{VLL}&=[\cC_{qu}^{(1)}]_{33ii}\delta_{ab}\delta_{cd}+[\cC_{qu}^{(8)}]_{33ii}T^A_{ab}T^A_{cd}.
\end{align}
The (spin and colour summed and averaged) interference terms with the SM read
\begin{align}
    2 \mathrm{Re}\left(\overline{\mathcal{A}_{VRR}^*\mathcal{A}_{SM}}\right)&= \frac{4 g_s^2}{9\hat{s}} [\cC_{qu}^{(8)}]_{ii33} \left(\hat{t}^2-m_t^2 (3\hat{t}+\hat{u}) +3m_t^4\right),\\
    2 \mathrm{Re}\left(\overline{\mathcal{A}_{VLL}^*\mathcal{A}_{SM}}\right)&= \frac{4 g_s^2}{9\hat{s}} [\cC_{qu}^{(8)}]_{33ii} \left(\hat{u}^2-m_t^2 (3\hat{u}+\hat{t}) +3m_t^4\right),
\end{align}
(where $\hat{t}=(p_1-p_3)^2$, $\hat{u}=(p_1-p_4)^2$ are the usual partonic Mandelstam variables) in agreement with previous literature~\cite{Zhang:2010dr,Degrande:2010kt}. The (spin and colour summed and averaged) NP$^2$ parts are then
\begin{align}
    |\overline{\mathcal{A}_{VRR}}|^2&=\left[\left([\cC_{qu}^{(1)}]_{33ii}\right)^2+\frac{2}{9}\left([\cC_{qu}^{(8)}]_{ii33}\right)^2\right] \left(\hat{t}^2-m_t^2 (3\hat{t}+\hat{u}) +3m_t^4\right),\\
    |\overline{\mathcal{A}_{VLL}}|^2&=\left[\left([\cC_{qu}^{(1)}]_{ii33}\right)^2+\frac{2}{9}\left([\cC_{qu}^{(8)}]_{33ii}\right)^2\right] \left(\hat{u}^2-m_t^2 (3\hat{u}+\hat{t}) +3m_t^4\right).
\end{align}
This gives a total partonic cross section for these vectorial operators of:
\begin{align}
    \hat{\sigma}&=\frac{\hat s}{48\pi}\left(1+\frac{2m_t^2}{\hat{s}} \right)\sqrt{1-\frac{4m_t^2}{\hat s}} \bigg[ \frac{4}{9}\frac{g_s^2}{\hat{s}}\left([\cC_{qu}^{(8)}]_{ii33}+[\cC_{qu}^{(8)}]_{33ii} \right) \nonumber \\
    &+\left([\cC_{qu}^{(1)}]_{ii33}\right)^2+\left([\cC_{qu}^{(1)}]_{33ii}\right)^2 +\frac{2}{9}\left([\cC_{qu}^{(8)}]_{ii33}\right)^2+\frac{2}{9}\left([\cC_{qu}^{(8)}]_{33ii}\right)^2\bigg].
\end{align}
Note that the $[\cC_{qu}^{(1,8)}]_{ii33}$ coefficients also contribute to $d_i \bar{d}_i\to t \bar t$ processes, unlike $[\cC_{qu}^{(1,8)}]_{33ii}$ and $[\cC_{qu}^{(1,8)}]_{i33i}$ which can only produce a $\bar t t$ final state from a $u_i \bar u_i$ initial state.

\subsection{Amplitudes and partonic cross section for $d_j\bar d_j \to t\bar t$ via $\cC_{quqd}^{(1,8)}$}
The relevant SMEFT tree-level amplitudes for the process $d_j^b (p_1) \bar d_j^a(p_2) \to t^c(p_3)\bar t^d(p_4)$ are
\begin{align}
    i \mathcal{A}_{S1}&=iC_{S1}\left[\bar{v}(p_2)^a P_Ru(p_1)^b\right]\left[\bar{u}(p_3)^c P_Rv(p_4)^d\right],\\
    i \mathcal{A}_{S2}&=iC_{S2}\left[\bar{v}(p_2)^a P_Rv(p_4)^d\right]\left[\bar{u}(p_3)^c P_Ru(p_1)^b\right],
\end{align}
where $a,b,c,d$ are colour indices and
\begin{align}
    C_{S1}&=[\cC_{quqd}^{(1)}]_{33jj}\,\delta_{ab}\delta_{cd}+\frac{1}{6}[\cC_{quqd}^{(8)}]_{33jj}(3\delta_{ad}\delta_{cb}-\delta_{ab}\delta_{cd}),\\
    C_{S2}&=[\cC_{quqd}^{(1)}]_{j33j}\,\delta_{ad}\delta_{cb}+\frac{1}{6}[\cC_{quqd}^{(8)}]_{j33j}(3\delta_{ab}\delta_{cd}-\delta_{ad}\delta_{cb}).
\end{align}
The (spin and colour summed and averaged) squared matrix elements are
\begin{align}
    |\overline{\mathcal{A}_{S1}}|^2&=\frac{1}{36}\hat{s}^2\left(1-\frac{2m_t^2}{\hat{s}} \right)|C_{S1}|^2,\\
    |\overline{\mathcal{A}_{S2}}|^2&=\frac{1}{36}\hat{t}^2\left(1-\frac{m_t^2}{\hat{t}} \right)|C_{S2}|^2\\
    2 \mathrm{Re}(\overline{\mathcal{A}_{S1}\mathcal{A}_{S2}^*})&=-\frac{2}{9}\hat{s}\hat{t}\,\mathrm{Re}\left( C_{S1}C_{S2}^*\right).
\end{align}
where
\begin{align}
|C_{S1}|^2&=9\left([\cC_{quqd}^{(1)}]_{33jj}\right)^2+2\left([\cC_{qu}^{(8)}]_{33jj}\right)^2,\\
|C_{S2}|^2&=9\left([\cC_{quqd}^{(1)}]_{j33j}\right)^2+2\left([\cC_{qu}^{(8)}]_{j33j}\right)^2,\\
C_{S1}C_{S2}^*&=[\cC_{quqd}^{(1)}]_{33jj}\left(3[\cC_{quqd}^{(1)}]_{j33j}^*+4[\cC_{quqd}^{(8)}]_{j33j}^*\right)\nonumber\\&+[\cC_{quqd}^{(8)}]_{33jj}\left(4[\cC_{quqd}^{(1)}]_{j33j}^*-\frac{2}{3}[\cC_{quqd}^{(8)}]_{j33j}^*\right).
\end{align}
Then the total partonic cross section is given by
\begin{align}
     \hat{\sigma}&=\frac{\hat s}{576\pi}\sqrt{1-\frac{4m_t^2}{\hat s}}\bigg[\left(1-\frac{2m_t^2}{\hat{s}} \right)\bigg(|C_{S1}|^2+\frac{16}{9}\mathrm{Re}\left( C_{S1}C_{S2}^*\right)\bigg)
     -\frac{1}{3}\left(1-\frac{5m_t^2}{2\hat{s}} \right)|C_{S2}|^2\bigg].
\end{align}

\acknowledgments
We are grateful to Dave Sutherland for comments on the manuscript. SR is partially supported by STFC grant ST/X000605/1. BS is supported by a Lord Kelvin Adam Smith scholarship from the University of Glasgow.

\bibliographystyle{JHEP}
\bibliography{refs}

\end{document}